\documentclass[12pt]{article}
\usepackage{amsmath}
\usepackage{color}
\usepackage{amssymb}
\usepackage{bm}

\makeatletter\@addtoreset{equation}{section}
\makeatother
\allowdisplaybreaks[1]
\begin{document}
\begin{titlepage}

\begin{flushright}
\phantom{preprint no.}
\end{flushright}
\vspace{0.5cm}
\begin{center}
{\Large \bf
Teukolsky-like equations in\\
\vspace{2mm}
a non-vacuum axisymmetric type D spacetime
}
\lineskip .75em
\vskip0.5cm
{\large Ya Guo${}^{1,\,2}$, Hiroaki Nakajima${}^{1}$ and Wenbin Lin${}^{1,\,2,\,*}$}
\vskip 2.5em
${}^{1}$ {\normalsize\it School of Mathematics and Physics, University of South China, \\ Hengyang, 421001, China\\
}
\vskip 1.0em
${}^{2}$ {\normalsize\it School of Physical Science and Technology, Southwest Jiaotong University, \\ Chengdu, 610031, China\\}
\vskip 1.0em
${}^{*}$ {\normalsize\it Email: lwb@usc.edu.cn\\}
\vskip 1.0em
\vskip 3.0em
\end{center}
\begin{abstract}
We study an axisymmetric metric satisfying the Petrov type D property with some additional ansatze, but without assuming the vacuum condition. We find that our metric in turn becomes conformal to the Kerr metric deformed by one function of the radial coordinate. We then study the gravitational-wave equations on this background metric in the case that the conformal factor is unity. We find that under an appropriate gauge condition,
the \textcolor{black}{homogeneous} wave equations admit the separation of the variables,
\textcolor{black}{which is also helpful for solving the nonhomogeneous equations.}
The \textcolor{black}{resultant ordinary differential} equation for the radial coordinate gives a natural extension of the Teukolsky equation.
\end{abstract}
\end{titlepage}

\section{Introduction}
The direct search of the gravitational waves is one of the most important topic in astrophysics
since the observation by LIGO and VIRGO~\cite{Abbott:2016blz}.
For the theoretical computation, the waves from the binary system \textcolor{black}{are} mainly studied, because it is much simpler than
the many-body system. In particular, for the case of the extreme mass-ratio inspiral (EMRI),
one can apply the black hole perturbation theory \cite{Mino:1997bx}, where the heavier object is regarded as the
(black hole) background, whereas the lighter object is regarded as the perturbation which appears as the source term of the wave equation.
For the background of the Kerr black hole, the wave equation propagating on it is derived by Teukolsky \cite{Teukolsky:1973ha},
and called Teukolsky equation. If the mass ratio of the binary is not so extreme,
one can apply the similar setting by simultaneously modifying the background and the perturbation by the (symmetric) mass-ratio.
This is called the effective one-boby (EOB) dynamics \cite{Buonanno:2000ef, Damour:2016gwp, Damour:2001tu}%
\footnote{Of course, one can also use the conventional post-Newtonian expansion method \cite{PoissonWill},
and the result from the EOB dynamics has to be compared with that. }.

Since the background of the EOB dynamics is deformed from the black hole background, it may not satisfy the vacuum condition.
Thus we need to study the wave equation on the non-vacuum background. Since it is however difficult in general,
here we would like to consider some simple cases. In the derivation of the Teukolsky equation, it plays a crucial role that
the Kerr background is classified as the Petrov type D \cite{Petrov}. Then we will assume that the background here also belongs to the type D,
but not assuming the vacuum condition. Here we note that in the case of the spherically symmetric background, it always belongs to the type D
\cite{Jing:2021ahx, Jing:2022vks, Guo:2023niy, Guo:2023hdn}.
\textcolor{black}{However, for the axisymmetric background, that is not always the case.}
%However, for the axisymmetric background, it does not always belong.
We will study a possible form of the background metric
under some ansatze which is introduced to avoid some technical complexities.

In the previous study on the spherically symmetric background, it has been found that the gravitational-wave equation in general
depends on the gauge \cite{Jing:2021ahx, Jing:2022vks, Guo:2023niy, Guo:2023hdn}, unless the background is vacuum,
which is unavoidable due to the gauge-dependence of the source term \cite{Guo:2023hdn}.
In the case of the axisymmetric background, one also has similar problems for the choice the gauge \cite{Jing:2023vzq}.
Moreover, the separation of the variables is quite nontrivial in the current case.
For instance, a certain gauge taken in \cite{Jing:2023vzq} does not allow the separation and needs the expansion in terms of the Kerr-like parameter,
which controls the axial deformation. We here consider the more convenient gauge such that the wave equation allows the separation of variables,
and the differential equation for the angular part in turn becomes that of the conventional spin-weighted spheroidal harmonics, as expected.

The remainder of this paper is organized as follows: in section \ref{BG}, we will show the background metric,
which in turn becomes conformal to the Kerr metric deformed by one function of the radial coordinate.
In section \ref{wave}, we will derive the gravitational-wave equation on the background where the conformal factor is unity.
Section \ref{summary} is devoted to the summary and discussion. In appendix \ref{sphesym}, we will summarize the reduction to the
spherically symmetric case. In appendix \ref{gaugecon}, we will discuss the gauge condition.

\section{Background metric and tetrads}\label{BG}

Here we will consider an axisymmetric \textcolor{black}{stationary} metric satisfying the Petrov type D condition using the Newman-Penrose formalism \cite{Newman:1961qr}.
We will also assume that the metric is asymptotically flat and some of the spin coefficients vanish,
which is to avoid the technical complexities and is for the later use. However we will not assume the vacuum condition.

We begin with the following form of the metric:
\begin{gather}
ds^{2}=A_{0}(dt-E_{1} d\varphi)^{2}-B(d\varphi-E_{2} dt)^{2}-C dr^{2}-D_{0} d\theta^{2},
\label{metric1}
\end{gather}
where $A_{0}$, $B$, $C$, $D_{0}$, $E_{1}$ and $E_{2}$ are the real functions of $r$ and $\theta$.
\eqref{metric1} is the most general form of the stationary axisymmetric metric.
We take the corresponding null tetrads as
\begin{align}
l&=dt-E_{1} d\varphi -\sqrt{\frac{C}{A_{0}}}dr\ ,
\notag\\
n&=\frac{A_{0}}{2}\left(dt-E_{1} d\varphi +\sqrt{\frac{C}{A_{0}}}dr\right),
\notag\\
m&=\frac{e^{i\xi}}{\sqrt{2}}\left[i\sqrt{B}(d\varphi-E_{2} dt)+\sqrt{D_{0}}d\theta\right],
\notag\\
\bar{m}&=\frac{e^{-i\xi}}{\sqrt{2}}\left[-i\sqrt{B}(d\varphi-E_{2} dt)+\sqrt{D_{0}}d\theta\right].
\label{tetrads1}
\end{align}
Here $\xi$ is a real function of $r$ and $\theta$. The null tetrads satisfy
\begin{gather}
ds^{2}=2ln-2m\bar{m}.
\end{gather}

We  now consider the type D condition. In order to avoid some technical complexities, we first require
$\kappa=\lambda=\sigma=\nu=0$, and then require the type D condition $\Psi_{0}=\Psi_{1}=\Psi_{3}=\Psi_{4}=0$ itself.
In the vacuum spacetime the former implies the latter, and vice versa, which is called the Goldberg-Sachs (GS) theorem \cite{GS} (for the type D).
Our strategy is inspired from the GS theorem and its non-vacuum extension \cite{KT, RS} (see also \cite{Stephani:2003tm} for instance).
Another reason to require $\kappa=\lambda=\sigma=\nu=0$ first is because we will also use it to derive the wave equation below.
%(see Appendix \ref{GS}).
The spin coefficients $\kappa$ and $\sigma$ are computed from \eqref{tetrads1} as
\begin{align}
\kappa&=-\frac{e^{i\xi}}{2\sqrt{2D_{0}}A_{0}}\left[
\frac{\partial_{\theta}C}{C}-\frac{\partial_{\theta}A_{0}}{A_{0}}+2\frac{E_{2}\partial_{\theta}E_{1}}{1-E_{1}E_{2}}
-2i\sqrt{\frac{A_{0}D_{0}}{BC}}\frac{\partial_{r}E_{1}}{1-E_{1}E_{2}}
\right],
\label{kappa}
\\
\sigma&=-\frac{e^{2i\xi}}{4\sqrt{A_{0}C}}\left[
\frac{\partial_{r}D_{0}}{D_{0}}-\frac{\partial_{r}B}{B}+2\frac{E_{1}\partial_{r}E_{2}}{1-E_{1}E_{2}}
+2i\sqrt{\frac{BC}{A_{0}D_{0}}}\frac{\partial_{\theta}E_{2}}{1-E_{1}E_{2}}
\right].
\label{sigma}
\end{align}
Then $\kappa=\sigma=0$ implies
\begin{align}
&\frac{\partial_{\theta}C}{C}-\frac{\partial_{\theta}A_{0}}{A_{0}}+2\frac{E_{2}\partial_{\theta}E_{1}}{1-E_{1}E_{2}}=0,
\label{cond1}
\\
&E_{1}=E_{1}(\theta),
\label{cond2}
\\
&\frac{\partial_{r}D_{0}}{D_{0}}-\frac{\partial_{r}B}{B}+2\frac{E_{1}\partial_{r}E_{2}}{1-E_{1}E_{2}}=0,
\label{cond3}
\\
&E_{2}=E_{2}(r),
\label{cond4}
\end{align}
where $\lambda=\nu=0$ also gives the same conditions. From \eqref{cond1} and \eqref{cond4}, we have
\begin{gather}
%\partial_{\theta}\left\{\ln\left[\frac{C}{A(1-E_{1}E_{2})^{2}}\right]\right\}=0,\quad \therefore
\frac{C}{A_{0}(1-E_{1}E_{2})^{2}}=F(r).
\label{rel1}
\end{gather}
In a similar way, from \eqref{cond2} and \eqref{cond3}, we have
\begin{gather}
%\partial_{r}\left\{\ln\left[\frac{D_{1}}{B(1-E_{1}E_{2})^{2}}\right]\right\}=0,\quad \therefore
\frac{D_{0}}{B(1-E_{1}E_{2})^{2}}=G(\theta).
\label{rel2}
\end{gather}
By the use of the above conditions and relations, one can find that $\Psi_{0}=\Psi_{4}=0$ is already satisfied.
$\Psi_{1}$ and $\Psi_{3}$ are computed as
\begin{align}
\Psi_{1}&=\frac{-ie^{i\xi}}{8\sqrt{2B}A_{0}(1-E_{1}E_{2})^{3}}(I-iJ),
\label{psi1}
\\
\Psi_{3}&=\frac{ie^{-i\xi}}{16\sqrt{2B}(1-E_{1}E_{2})^{3}}(I-iJ),
\label{psi3}
\end{align}
where $I$ and $J$ are given by
\begin{align}
%I&=
%\frac{2}{F(r)}\partial_{r}\left(\frac{B(r,\theta)}{A_{0}(r,\theta)}E'_{2}(r)\right)
%+\frac{2}{G(\theta)}\partial_{\theta}\left(\frac{A_{0}(r,\theta)}{B(r,\theta)}E'_{1}(\theta)\right)
%\notag\\
%&\qquad {}+(1-E_{1}E_{2})^{2}\frac{B(r,\theta)}{A_{0}(r,\theta)}E'_{2}(r)\partial_{r}
%\left[\frac{1}{F(r)(1-E_{1}(r)E_{2}(\theta))^{2}}\right]
%\notag\\
%&\qquad {}+(1-E_{1}E_{2})^{2}\frac{A_{0}(r,\theta)}{B(r,\theta)}E'_{1}(\theta)\partial_{\theta}
%\left[\frac{1}{G(\theta)(1-E_{1}(r)E_{2}(\theta))^{2}}\right],
I&=
\frac{2}{F}\partial_{r}\left(\frac{B}{A_{0}}E'_{2}(r)\right)+\frac{2}{G}\partial_{\theta}\left(\frac{A_{0}}{B}E'_{1}(\theta)\right)
\notag\\
&\qquad {}+(1-E_{1}E_{2})^{2}\frac{B}{A_{0}}E'_{2}(r)\partial_{r}\left[\frac{1}{F(1-E_{1}E_{2})^{2}}\right]
\notag\\
&\qquad {}+(1-E_{1}E_{2})^{2}\frac{A_{0}}{B}E'_{1}(\theta)\partial_{\theta}\left[\frac{1}{G(1-E_{1}E_{2})^{2}}\right],
\label{I}
\\
J&=\frac{2(1-E_{1}E_{2})}{\sqrt{FG}}\left[
\frac{\partial_{r}\partial_{\theta}B}{B}-\frac{\partial_{r}B\partial_{\theta}B}{B^{2}}
-\frac{\partial_{r}\partial_{\theta}A_{0}}{A_{0}}+\frac{\partial_{r}A_{0}\partial_{\theta}A_{0}}{A_{0}^{2}}
\right].
\label{J}
\end{align}
\textcolor{black}{Here} the prime denotes the ordinary derivative of the functions with respect to their arguments.
$\Psi_{1}=\Psi_{3}=0$ implies $I=J=0$. The condition $J=0$ can be integrated as
\begin{gather}
\frac{B(r,\theta)}{A_{0}(r,\theta)}=H(r)M(\theta).
\label{rel3}
\end{gather}
By substituting \eqref{rel1}, \eqref{rel2} and \eqref{rel3}, the metric \eqref{metric1} becomes
\begin{align}
ds^{2}&=A_{0}\Bigl[(dt-E_{1} d\varphi)^{2}-HM(d\varphi-E_{2} dt)^{2}
\notag\\
&\qquad\qquad
{}-F(1-E_{1}E_{2})^{2} dr^{2}-HMG(1-E_{1}E_{2})^{2} d\theta^{2}\Bigr].
\label{metric2}
\end{align}

In order to simplify the condition $I=0$,
we use the condition such that the metric \eqref{metric1} is asymptotically flat, which
gives the behavior of the functions at large $r$ as
\begin{align}
A_{0}(r,\theta)&\sim 1,
\label{af1}
\\
E_{1}(\theta)-E_{2}(r)H(r)M(\theta)&\sim 0,
\label{af2}
\\
H(r)M(\theta)-E_{1}^{2}(\theta)&\sim r^{2}\sin^{2}\theta,
\label{af3}
\\
F(r)[1-E_{1}(\theta)E_{2}(r)]^{2}&\sim 1,
\label{af4}
\\
H(r)M(\theta)G(\theta)[1-E_{1}(\theta)E_{2}(r)]^{2}&\sim r^{2}.
\label{af5}
\end{align}
From \eqref{af2}, the product $E_{2}(r)H(r)$ has to behave as a constant at large $r$. Thus we define a constant $a$ as
\begin{gather}
a=\lim_{r\to\infty}E_{2}(r)H(r).
\end{gather}
We will see later that $a$ plays a similar role of the Kerr parameter.
Other conditions fix the functions of $\theta$ explicitly and behaviors of the functions of $r$ at large $r$ as
\begin{align}
E_{1}(\theta)&=a\sin^{2}\theta, & G(\theta)&=\frac{1}{\sin^{2}\theta}, & M(\theta)&=\sin^{2}\theta,
\label{theta}
\\
E_{2}(r)&\sim \frac{a}{r^{2}}, & F(r)&\sim 1, & H(r)&\sim r^{2}.
\label{r}
\end{align}
From \eqref{theta} and \eqref{r}, $I$ in \eqref{I} is evaluated as
\begin{align}
I&=-4a+8a^{2}E_{2}+P'E'_{2}+2PE''_{2}
\notag\\
&\qquad {}+a\sin^{2}\theta\left[-4aE_{2}-P'E_{2}E'_{2}-2PE_{2}E''_{2}+2P(E'_{2})^{2}\right],
\end{align}
where $P(r)=H^{2}(r)/F(r)$.
Then the condition $I=0$ is split into two ordinary differential equations with respect to the argument $r$ as
\begin{align}
-4a+8a^{2}E_{2}+P'E'_{2}+2PE''_{2}&=0,
\label{diff1}
\\
-4aE_{2}-P'E_{2}E'_{2}-2PE_{2}E''_{2}+2P(E'_{2})^{2}&=0.
\label{diff2}
\end{align}
By eliminating $P'$ from two equations, we obtain
\begin{gather}
P=\frac{4aE_{2}-4a^{2}E_{2}^{2}}{(E'_{2})^{2}}.
\label{rel4}
\end{gather}
One can confirm that \eqref{rel4} satisfies both \eqref{diff1} and \eqref{diff2}.
By substituting \eqref{theta} and \eqref{rel4}, the metric \eqref{metric2} is
\begin{align}
ds^{2}&=A_{0}\Biggl[(dt-a\sin^{2}\theta d\varphi)^{2}-H\sin^{2}\theta(d\varphi-E_{2} dt)^{2}
\notag\\
&\qquad\qquad
{}-\frac{H^{2}(E'_{2})^{2}}{4aE_{2}-4a^{2}E_{2}^{2}}(1-a\sin^{2}\theta E_{2})^{2} dr^{2}
-H(1-a\sin^{2}\theta E_{2})^{2} d\theta^{2}\Biggr].
\label{metric3}
\end{align}

We define the function $R(r)$ as
\begin{gather}
R(r)=\sqrt{\frac{a}{E_{2}(r)}-a^{2}},
\end{gather}
and then \eqref{metric3} is rewritten as
\begin{align}
ds^{2}&=A_{0}\Biggl[(dt-a\sin^{2}\theta d\varphi)^{2}-H\sin^{2}\theta\left(d\varphi-\frac{a}{R^{2}+a^{2}} dt\right)^{2}
\notag\\
&\qquad\qquad
{}-\frac{H^{2}(R')^{2}}{(R^{2}+a^{2})^{2}}\left(\frac{\Sigma}{R^{2}+a^{2}}\right)^{2} dr^{2}
-H\left(\frac{\Sigma}{R^{2}+a^{2}}\right)^{2} d\theta^{2}\Biggr].
\label{metric4}
\end{align}
where $\Sigma$ is defined by
\begin{gather}
\Sigma=R(r)^{2}+a^{2}\cos^{2}\theta.
\end{gather}
The existence of the factor $(R')^{2}dr^{2}=dR^{2}$ \textcolor{black}{implies} that one can use $R$ as the radial coordinate,
which is equivalent to take the gauge
\begin{gather}
R(r)=r.
\label{gauge}
\end{gather}
In this gauge \eqref{metric4} becomes
\begin{align}
ds^{2}&=A_{0}\Biggl[(dt-a\sin^{2}\theta d\varphi)^{2}-\frac{H^{2}\Sigma^{2}}{(r^{2}+a^{2})^{4}} dr^{2}
\notag\\
&\qquad\qquad\quad {}
-\frac{H\Sigma^{2}}{(r^{2}+a^{2})^{2}} d\theta^{2}-H\sin^{2}\theta\left(d\varphi-\frac{a}{r^{2}+a^{2}} dt\right)^{2}\Biggr].
\label{metric5}
\end{align}
Finally, by defining
\begin{gather}
L(r)=\frac{(r^{2}+a^{2})^{2}}{H(r)}, \quad A(r,\theta)=A_{0}(r,\theta)\frac{\Sigma}{L(r)},
\end{gather}
we obtain the metric as
\begin{align}
ds^{2}&=A(r,\theta)\Biggl[\frac{L(r)}{\Sigma}(dt-a\sin^{2}\theta d\varphi)^{2}-\frac{\Sigma}{L(r)}dr^{2}
\notag\\
&\qquad\qquad\quad {}
-\Sigma d\theta^{2}-\frac{(r^{2}+a^{2})^{2}\sin^{2}\theta}{\Sigma}\left(d\varphi-\frac{a}{r^{2}+a^{2}} dt\right)^{2}\Biggr],
\label{metric7}
\end{align}One can find that \eqref{metric7} is the Kerr metric deformed by the function $L(r)$
and by the conformal factor $A(r,\theta)$. When we restrict these two functions as
\begin{gather}
A(r,\theta)=1,\quad L(r)=r^{2}-2Mr+a^{2},
\end{gather}
the metric \eqref{metric7} reduces to the Kerr metric with the mass parameter $M$ and the Kerr parameter $a$
in the Boyer-Lindquist coordinate, where the choice of the coordinate is from the gauge condition \eqref{gauge}.
On the other hand, when we consider the case $A(r,\theta)=A(r)$ and $a=0$,
\eqref{metric7} reduces to the general form of the spherically symmetric metric in a particular gauge (see appendix \ref{sphesym}).

The null tetrads \eqref{tetrads1} become
\begin{align}
l&=dt-a\sin^{2}\theta d\varphi -\frac{\Sigma}{L}dr\ ,
\notag\\
n&=\frac{A}{2}\left[\frac{L}{\Sigma}(dt-a\sin^{2}\theta d\varphi) +dr\right],
\notag\\
m&=\frac{1}{\varrho}\sqrt{\frac{A}{2}}\left[-i(r^{2}+a^{2})\sin\theta d\varphi+ia\sin\theta dt-\Sigma d\theta\right],
\notag\\
\bar{m}&=\frac{1}{\bar{\varrho}}\sqrt{\frac{A}{2}}\left[i(r^{2}+a^{2})\sin\theta d\varphi- ia\sin\theta dt-\Sigma d\theta\right],
\label{tetrads2}
\end{align}
where $\varrho=r+ia\cos\theta$ and the phase $\xi$ is chosen as
\begin{gather}
e^{-i\xi}=-\frac{\varrho}{\sqrt{\Sigma}}=-\frac{\varrho}{|\varrho|}.
\end{gather}
From the tetrad basis \eqref{tetrads2}, the spin coefficients are computed as
\begin{gather}
%\kappa=\lambda=\sigma=\nu=\epsilon=0,
%\\
%\rho=-\frac{1}{A\bar{\varrho}}-\frac{\partial_{r}A}{2A^{2}},
%\quad
%\tau=-\frac{ia\sin\theta}{\sqrt{2A}\Sigma}-\frac{\partial_{\theta}A}{2\sqrt{2A^{3}}\varrho},
%\quad
%\mu=-\frac{L}{2\bar{\varrho}\Sigma}-\frac{L\partial_{r}A}{4A\Sigma},
%\\
%\gamma=\mu+\frac{L'}{4\Sigma}+\frac{L \partial_{r}A}{2A \Sigma},
%\quad
%\pi=\frac{ia\sin\theta}{\sqrt{2A}\bar{\varrho}^{2}}+\frac{\partial_{\theta}A}{2\sqrt{2A^{3}}\bar{\varrho}},
%\quad
%\beta=\frac{\cot\theta}{2\sqrt{2A}\varrho},
%\quad \\
%\alpha=\pi-\bar{\beta}-\frac{\partial_{\theta}A}{\sqrt{2A^{3}}\bar{\varrho}}.
%\alpha=\frac{ia\sin\theta}{\sqrt{2A}\bar{\varrho}^{2}}-\frac{\cot\theta}{2\sqrt{2A}\bar{\varrho}}-\frac{\partial_{\theta}A}{2\sqrt{2A^{3}}\bar{\varrho}}
\kappa=\lambda=\sigma=\nu=\epsilon=0,
\\
\mu=-\frac{L}{2\bar{\varrho}\Sigma}-\frac{L\partial_{r}A}{4A\Sigma},
\quad
\tau=-\frac{ia\sin\theta}{\sqrt{2A}\Sigma}-\frac{\partial_{\theta}A}{2\sqrt{2A^{3}}\varrho},
\quad
\pi=\frac{ia\sin\theta}{\sqrt{2A}\bar{\varrho}^{2}}+\frac{\partial_{\theta}A}{2\sqrt{2A^{3}}\bar{\varrho}},
\\
\rho=-\frac{1}{A\bar{\varrho}}-\frac{\partial_{r}A}{2A^{2}},
\quad
\gamma=\mu+\frac{L'}{4\Sigma}+\frac{L \partial_{r}A}{2A \Sigma},
\quad
\beta=\frac{\cot\theta}{2\sqrt{2A}\varrho},
\quad
\alpha=\pi-\bar{\beta}-\frac{\partial_{\theta}A}{\sqrt{2A^{3}}\bar{\varrho}}.
\end{gather}
In a similar way, the Weyl scalars are
\begin{gather}
\Psi_{0}=\Psi_{1}=\Psi_{3}=\Psi_{4}=0,
\\
%\begin{aligned}
\Psi_{2}=\frac{1}{A\Sigma}\left[
-\frac{1}{6}-\frac{a^{2}\sin^{2}\theta}{\bar{\varrho}^{2}}-\frac{ia\cos\theta}{\bar{\varrho}}+\frac{L}{\bar{\varrho}^{2}}
-\frac{L'}{2\bar{\varrho}}+\frac{L''}{12}
\right].
%\end{aligned}
\end{gather}
Finally the components of the Ricci tensor are
\begin{gather}
\Phi_{00}=\frac{3(\partial_{r}A)^{2}}{4A^{4}}-\frac{\partial_{r}^{2}A}{2A^{3}},
\\
\Phi_{01}=\bar{\Phi}_{10}=\frac{\partial_{\theta}A}{2\sqrt{2A^{5}}\varrho^{2}}
-\frac{ia\sin\theta\partial_{r}A}{2\sqrt{2A^{5}}\varrho^{2}}
+\frac{3\partial_{r}A\partial_{\theta}A}{4\sqrt{2A^{7}}\varrho}
-\frac{\partial_{r}\partial_{\theta}A}{2\sqrt{2A^{5}}\varrho},
\\
\begin{aligned}
\Phi_{11}&=\frac{1}{4A\Sigma}
\left[\frac{2r^{2}-2a^{2}+2L-2rL'}{\Sigma}+\frac{L''}{2}-1
-\left(\frac{\cot\theta}{2}+\frac{a^{2}\sin 2\theta}{\Sigma}\right)\frac{\partial_{\theta}A}{A}\right.
\\
&\quad {}+\left.
\frac{3}{4}\left(\frac{\partial_{\theta}A}{A}\right)^{2}
+\left(\frac{L'}{2}-\frac{2rL}{\Sigma}\right)\frac{\partial_{r}A}{A}
-\frac{3}{4}L\left(\frac{\partial_{r}A}{A}\right)^{2}
-\frac{\partial_{\theta}^{2}A}{2A}+\frac{L\partial_{r}^{2}A}{2A}
\right],
\end{aligned}
\\
\Phi_{02}=\bar{\Phi}_{20}=\frac{3(\partial_{\theta}A)^{2}}{8A^{3}\varrho^{2}}
+\frac{\cot\theta\partial_{\theta}A}{4A^{2}\varrho^{2}}
-\frac{\partial_{\theta}^{2}A}{4A^{2}\varrho^{2}},
\\
\Phi_{22}=\frac{3L^{2}(\partial_{r}A)^{2}}{16A^{2}\Sigma^{2}}-\frac{L^{2}\partial_{r}^{2}A}{8A\Sigma^{2}},
\\
\Phi_{12}=\bar{\Phi}_{21}=-\frac{3\bar{\varrho}L\partial_{r}A \partial_{\theta}A}{8\sqrt{2A^{5}}\Sigma^{2}}
-\frac{L\partial_{\theta}A}{4\sqrt{2A^{3}}\Sigma^{2}}
-\frac{ia\sin\theta L\partial_{r}A}{4\sqrt{2A^{3}}\Sigma^{2}}
+\frac{\bar{\varrho}L\partial_{r}\partial_{\theta}A}{4\sqrt{2A^{3}}\Sigma^{2}},
\\
\begin{aligned}
\Lambda&=
\frac{1}{4A\Sigma}\left[\frac{1}{3}-\frac{L''}{6}-\frac{\cot\theta\partial_{\theta}A}{2A}+\frac{1}{4}\left(\frac{\partial_{\theta}A}{A}\right)^{2}
\right.
\\
&\qquad\qquad {}-\left.\frac{L'\partial_{r}A}{2A}+\frac{L}{4}\left(\frac{\partial_{r}A}{A}\right)^{2}
-\frac{\partial_{\theta}^{2}A}{2A}-\frac{L\partial_{r}^{2}A}{2A}
\right].
\end{aligned}
\end{gather}
Note that in the case of $A(r,\theta)=1$, the spin coefficients, the Weyl scalars and the components of the Ricci tensor are simplified as
\begin{gather}
\kappa=\lambda=\sigma=\nu=\epsilon=0,
\\
\rho=-\frac{1}{\bar{\varrho}},
\quad
\tau=-\frac{ia\sin\theta}{\sqrt{2}\Sigma},
\quad
\mu=-\frac{L}{2\bar{\varrho}\Sigma},
\\
\gamma=\mu+\frac{L'}{4\Sigma},
\quad
\pi=\frac{ia\sin\theta}{\sqrt{2}\bar{\varrho}^{2}},
\quad
\beta=\frac{\cot\theta}{2\sqrt{2}\varrho},
\quad
\alpha=\pi-\bar{\beta},
\\
\Psi_{0}=\Psi_{1}=\Psi_{3}=\Psi_{4}=0,
\\
%\begin{aligned}
\Psi_{2}=\frac{1}{\Sigma}\left[
-\frac{1}{6}-\frac{a^{2}\sin^{2}\theta}{\bar{\varrho}^{2}}-\frac{ia\cos\theta}{\bar{\varrho}}+\frac{L}{\bar{\varrho}^{2}}
-\frac{L'}{2\bar{\varrho}}+\frac{L''}{12}
\right],
%\end{aligned}
\\
\Phi_{00}=\Phi_{01}=\Phi_{10}=\Phi_{02}=\Phi_{20}=\Phi_{12}=\Phi_{21}=0,
\label{simple}
\\
\Phi_{11}=\frac{1}{4\Sigma}
\left[\frac{2r^{2}-2a^{2}+2L-2rL'}{\Sigma}+\frac{L''}{2}-1\right],
\label{simple1}
\\
\Lambda=\frac{2-L''}{24\Sigma}.
\label{simple2}
\end{gather}
In particular, some of the components of the Ricci tensor vanish as \eqref{simple}, which gives the great simplification when we will study
the gravitational-wave equation on this background in next section. We also note that in the case of $A(r,\theta)=1$,
our background \eqref{metric7} coincides the one studied in \cite{Jing:2023vzq},
which is the special case of the axisymmetric EOB background in \cite{Damour:2001tu}.
In \cite{Damour:2001tu}, the function $L(r)$ in front of $(dt-a\sin^{2}\theta d\varphi)^{2}$ and that in front of $dr^{2}$
are taken to be different. However in that case, the type D condition cannot be \textcolor{black}{satisfied}.

%\section{Wave equation for perturbed Weyl scalars}
\section{Gravitational-wave equation}\label{wave}

\textcolor{black}{In the previous section}, we have shown that the metric \eqref{metric7} is classified as
the non-vacuum Petrov type D background, which is useful to derive the
gravitational-wave equation for the perturbed Weyl scalars using the Newman-Penrose formalism,
as in the case of the Teukolsky equation \cite{Teukolsky:1973ha} on the Schwarzschild and the Kerr background.
For simplicity, hereafter we will focus on the case $A(r,\theta)=1$, such that among the components of the Ricci tensor,
only $\Phi_{11}$ and $\Lambda$ are nonvanishing as \eqref{simple}--\eqref{simple2}.
The case with the general $A(r,\theta)$ will be left as the future work.
We note that in the case of the spherically symmetric background, the wave equation with the general conformal factor
can be derived \textcolor{black}{\cite{Guo:2023niy,Guo:2023hdn}}, where the different radial coordinate is used.
For the comparison of the radial coordinate, see appendix \ref{sphesym}.

We begin with the following equations in Newman-Penrose formalism:
\begin{align}
&(\delta+4\beta-\tau)\Psi_{4}-(\Delta+4\mu+2\gamma)\Psi_{3}+3\nu\Psi_{2} \notag\\
&\qquad =(\bar{\delta}-\bar{\tau}+2\bar{\beta}+2\alpha)\Phi_{22}-(\Delta+2\gamma+2\bar{\mu})\Phi_{21}
-2\lambda\Phi_{12}+2\nu\Phi_{11}+\bar{\nu}\Phi_{20}, \label{eq1}\\
&(D+4\epsilon-\rho)\Psi_{4}-(\bar{\delta}+4\pi+2\alpha)\Psi_{3}+3\lambda\Psi_{2} \notag\\
&\qquad =(\bar{\delta}-2\bar{\tau}+2\alpha)\Phi_{21}-(\Delta+2\gamma-2\bar{\gamma}+\bar{\mu})\Phi_{20}
+\bar{\sigma}\Phi_{22}-2\lambda\Phi_{11}+2\nu\Phi_{10}, \label{eq2}\\
&(\Delta+\mu+\bar{\mu}+3\gamma-\bar{\gamma})\lambda-(\bar{\delta}+\pi-\bar{\tau}+\bar{\beta}+3\alpha)\nu+\Psi_{4}=0.
\label{eq3}
\end{align}
We split all the quantities in the above into the background part $(A)$ and the perturbation part $(B)$,
for instance, $\Psi_{4}=\Psi_{4}^{A}+\Psi_{4}^{B}$, etc.
The background part of eqs.\eqref{eq1}--\eqref{eq3} are shown to be satisfied,
and the part of the first-order perturbation becomes
\begin{align}
&(\delta+4\beta-\tau)^{A}\Psi_{4}^{B}-(\Delta+4\mu+2\gamma)^{A}\Psi_{3}^{B}+3\nu^{B}\Psi_{2}^{A} \notag\\
&\qquad =(\bar{\delta}-\bar{\tau}+2\bar{\beta}+2\alpha)^{A}\Phi_{22}^{B}
-(\Delta+2\gamma+2\bar{\mu})^{A}\Phi_{21}^{B}+2\nu^{B}\Phi_{11}^{A},
\label{eq01}
\\
&(D+4\epsilon-\rho)^{A}\Psi_{4}^{B}-(\bar{\delta}+4\pi+2\alpha)^{A}\Psi_{3}^{B}+3\lambda^{B}\Psi_{2}^{A} \notag\\
&\qquad =(\bar{\delta}-2\bar{\tau}+2\alpha)^{A}\Phi_{21}^{B}-(\Delta+2\gamma-2\bar{\gamma}+\bar{\mu})^{A}\Phi_{20}^{B}
-2\lambda^{B}\Phi_{11}^{A},
\label{eq02}
\\
&(\Delta+\mu+\bar{\mu}+3\gamma-\bar{\gamma})^{A}\lambda^{B}-(\bar{\delta}+\pi-\bar{\tau}+\bar{\beta}+3\alpha)^{A}\nu^{B}
+\Psi_{4}^{B}=0,
\label{eq03}
\end{align}
where the superscript $A$ $(B)$ on the parentheses denotes that all the quantities
and the operators inside the parentheses are in the background (perturbation). The quantities in previous section are
regarded as the background ones, and hence they will appear in this section with the superscript $A$.

We will obtain the wave equation for $\Psi_{4}^{B}$ in a similar way with the method
used to derive the Teukolsky equation \cite{Teukolsky:1973ha}. First we note that the following commutation relation of the
differential operators holds not only in \textcolor{black}{the} vacuum but also in the current background:
\begin{align}
&\left[\Delta+(p+1)\gamma-\bar{\gamma}-q\mu+\bar{\mu}\right]^{A}(\bar{\delta}+p\alpha-q\pi)^{A} \notag\\
&\qquad -\left[\bar{\delta}+(p+1)\alpha+\bar{\beta}-\bar{\tau}-q\pi\right]^{A}(\Delta+p\gamma-q\mu)^{A} \notag\\
&=\nu^{A}D^{A}-\lambda^{A}\delta^{A}-p\left[(\beta+\tau)\lambda-(\rho+\epsilon)\nu+\Psi_{3}\right]^{A} \notag\\
&\qquad +q\left[-D\nu+\delta\lambda+(\bar{\pi}+\tau+3\beta-\bar{\alpha})\lambda
-(3\epsilon+\bar{\epsilon}+\rho-\bar{\rho})\nu+2\Psi_{3}\right]^{A} \notag\\
&=0, \label{com}
\end{align}
where $p$ and $q$ are arbitrary constants and we have just used%
\footnote{%Here $\lambda^{A}=\nu^{A}=0$ is important.
$\kappa^{A}=\sigma^{A}=0$ will also be used for a similar calculation of the
wave equation for $\Psi_{0}^{B}$. }
$\lambda^{A}=\nu^{A}=\Psi_{3}^{A}=0$.
We operate $(\Delta+3\gamma-\bar{\gamma}+4\mu+\bar{\mu})^{A}$ to \eqref{eq02}
and $(\bar{\delta}+3\alpha+\bar{\beta}-\bar{\tau}+4\pi)^{A}$ to \eqref{eq01}, and then subtract one equation
from the other. The terms with $\Psi_{3}^{B}$ cancel by \eqref{com} with $p=2$, $q=-4$
and the remaining becomes
\begin{align}
&\left[(\Delta+3\gamma-\bar{\gamma}+4\mu+\bar{\mu})(D+4\epsilon-\rho)
-(\bar{\delta}+3\alpha+\bar{\beta}-\bar{\tau}+4\pi)(\delta+4\beta-\tau)\right]^{A}\Psi_{4}^{B} \notag\\
&\qquad {}+(3\Psi_{2}+2\Phi_{11})^{A}(\Delta+3\gamma-\bar{\gamma}+4\mu+\bar{\mu})^{A}\lambda^{B} \notag\\
&\qquad {}-(3\Psi_{2}-2\Phi_{11})^{A}(\bar{\delta}+3\alpha+\bar{\beta}-\bar{\tau}+4\pi)^{A}\nu^{B} \notag\\
&=T_{4} -\lambda^{B}\Delta^{A}(3\Psi_{2}+2\Phi_{11})^{A}+\nu^{B}\bar{\delta}^{A}(3\Psi_{2}-2\Phi_{11})^{A}
\label{eq04}
\end{align}
where $T_{4}$ is defined by
\begin{align}
T_{4}&=(\Delta+3\gamma-\bar{\gamma}+4\mu+\bar{\mu})^{A}
\left[(\bar{\delta}-2\bar{\tau}+2\alpha)^{A}\Phi_{21}^{B}
-(\Delta+2\gamma-2\bar{\gamma}+\bar{\mu})^{A}\Phi_{20}^{B}\right]
\notag\\
&\quad {}-(\bar{\delta}+3\alpha+\bar{\beta}-\bar{\tau}+4\pi)^{A}
\left[(\bar{\delta}-\bar{\tau}+2\bar{\beta}+2\alpha)^{A}\Phi_{22}^{B}
-(\Delta+2\gamma+2\bar{\mu})^{A}\Phi_{21}^{B}\right].
\end{align}
In the right hand side, we use
\begin{align}
\Delta^{A}(3\Psi_{2}-2\Phi_{11})^{A}&=-3\mu^{A}(3\Psi_{2}+2\Phi_{11})^{A}+4(\mu+\bar{\mu})^{A}\Phi_{11}^{A},
\\
\bar{\delta}^{A}(3\Psi_{2}+2\Phi_{11})^{A}&=-3\pi^{A}(3\Psi_{2}-2\Phi_{11})^{A}-4(\pi-\bar{\tau})^{A}\Phi_{11}^{A}.
\end{align}
and then substituting the above into \eqref{eq04}, we have
\begin{align}
&\left[(\Delta+3\gamma-\bar{\gamma}+4\mu+\bar{\mu})(D+4\epsilon-\rho)
-(\bar{\delta}+3\alpha+\bar{\beta}-\bar{\tau}+4\pi)(\delta+4\beta-\tau)-3\Psi_{2}\right]^{A}\Psi_{4}^{B} \notag\\
&\qquad {}+2\Phi_{11}^{A}(\Delta+3\gamma-\bar{\gamma}+\mu+\bar{\mu})^{A}\lambda^{B}
+2\Phi_{11}^{A}(\bar{\delta}+3\alpha+\bar{\beta}-\bar{\tau}+\pi)^{A}\nu^{B} \notag\\
&=T_{4} -4\lambda^{B}[(\Delta+\mu+\bar{\mu})\Phi_{11}]^{A}-4\nu^{B}[(\bar{\delta}+\pi-\bar{\tau})\Phi_{11}]^{A},
\label{eq05}
\end{align}
where we have used \eqref{eq03}.
In the case of vacuum, \eqref{eq05} gives the decoupled equation for $\Psi_{4}^{B}$ \textcolor{black}{since $\Phi_{11}^{A}=0$}.
\textcolor{black}{Here the term ``decoupled equation'' means that the equation does not contain the other perturbed Weyl scalars
and the perturbed spin coefficients.
However} in the present case,
$\nu^{B}$ and $\lambda^{B}$ are still remaining. Hence by applying appropriate gauge conditions, one has to eliminate them.

One possible choice is to take the gauge $\Psi_{3}^{B}=0$ \cite{Jing:2023vzq},
under which both $\lambda^{B}$ and $\nu^{B}$ can be expressed in terms of $\Psi_{4}^{B}$ from \eqref{eq01} and \eqref{eq02}.
However the resulting wave equation becomes complicated, and then we will consider other gauges.
Since the relative sign in the second line in \eqref{eq05} is positive, one cannot rewrite the second line in terms of $\Psi_{4}^{B}$
using \eqref{eq03} completely, but instead one can eliminate either $\nu^{B}$ or $\lambda^{B}$. When $\nu^{B}$ is eliminated, \eqref{eq05} becomes
\begin{align}
&\left[(\Delta+3\gamma-\bar{\gamma}+4\mu+\bar{\mu})(D+4\epsilon-\rho)\right. \notag\\
&\qquad {}-\left.(\bar{\delta}+3\alpha+\bar{\beta}-\bar{\tau}+4\pi)(\delta+4\beta-\tau)-3\Psi_{2}+2\Phi_{11}\right]^{A}\Psi_{4}^{B} \notag\\
&=T_{4} -4(\Delta+3\gamma-\bar{\gamma}+2\mu+2\bar{\mu})^{A}(\Phi_{11}^{A}\lambda^{B})-4\nu^{B}[(\bar{\delta}+\pi-\bar{\tau})\Phi_{11}]^{A}.
\label{eq06}
\end{align}
Then under the gauge condition
\begin{gather}
\Omega_{1}\equiv(\Delta+3\gamma-\bar{\gamma}+2\mu+2\bar{\mu})^{A}(\Phi_{11}^{A}\lambda^{B})+\nu^{B}[(\bar{\delta}+\pi-\bar{\tau})\Phi_{11}]^{A}=0,
\label{gauge01}
\end{gather}
we obtain the decoupled equation for $\Psi_{4}^{B}$ as
\begin{align}
&\left[(\Delta+3\gamma-\bar{\gamma}+4\mu+\bar{\mu})(D+4\epsilon-\rho)\right. \notag\\
&\qquad {}-\left.(\bar{\delta}+3\alpha+\bar{\beta}-\bar{\tau}+4\pi)(\delta+4\beta-\tau)-3\Psi_{2}+2\Phi_{11}\right]^{A}\Psi_{4}^{B}=T_{4}.
\label{waveeq01}
\end{align}
In the case of spherically symmetric background this gauge is simple, because the terms with $\nu^{B}$ in the gauge condition \eqref{gauge01}
vanish due to
\begin{gather}
(\bar{\delta}+\pi-\bar{\tau})^{A}\Phi_{11}^{A}=0,
\label{sphesymeq}
\end{gather}
and then \eqref{gauge01} is satisfied just by setting $\lambda^{B}=0$, which has been used in the previous studies.
In the present case, \eqref{sphesymeq} does not hold generically unless the vacuum case,
but instead, one can solve \eqref{gauge01} in terms of $\nu^{B}$ as
\begin{gather}
\nu^{B}= -\frac{(\Delta+3\gamma-\bar{\gamma}+2\mu+2\bar{\mu})^{A}(\Phi_{11}^{A}\lambda^{B})}{(\bar{\delta}+\pi-\bar{\tau})^{A}\Phi_{11}^{A}}.
\label{gauge001}
\end{gather}
When we eliminate $\lambda^{B}$ from the second line in \eqref{eq05}, we have
\begin{align}
&\left[(\Delta+3\gamma-\bar{\gamma}+4\mu+\bar{\mu})(D+4\epsilon-\rho)\right. \notag\\
&\qquad {}-\left.(\bar{\delta}+3\alpha+\bar{\beta}-\bar{\tau}+4\pi)(\delta+4\beta-\tau)-3\Psi_{2}-2\Phi_{11}\right]^{A}\Psi_{4}^{B} \notag\\
&=T_{4} -4\lambda^{B}[(\Delta+\mu+\bar{\mu})\Phi_{11}]^{A}-4(\bar{\delta}+3\alpha+\bar{\beta}-2\bar{\tau}+2\pi)^{A}(\Phi_{11}^{A}\nu^{B}).
\label{eq07}
\end{align}
Then under the gauge condition
\begin{gather}
 \Omega_{2}\equiv\lambda^{B}[(\Delta+\mu+\bar{\mu})\Phi_{11}]^{A}+(\bar{\delta}+3\alpha+\bar{\beta}-2\bar{\tau}+2\pi)^{A}(\Phi_{11}^{A}\nu^{B})=0,
\label{gauge02}
\end{gather}
or equivalently\footnote{One can find that the denominator of \eqref{gauge002} is generically nonvanishing unless the vacuum case. }
\begin{gather}
\lambda^{B}= -\frac{(\bar{\delta}+3\alpha+\bar{\beta}-2\bar{\tau}+2\pi)^{A}(\Phi_{11}^{A}\nu^{B})}{(\Delta+\mu+\bar{\mu})^{A}\Phi_{11}^{A}},
\label{gauge002}
\end{gather}
we obtain the decoupled equation as
\begin{align}
&\left[(\Delta+3\gamma-\bar{\gamma}+4\mu+\bar{\mu})(D+4\epsilon-\rho)\right. \notag\\
&\qquad {}-\left.(\bar{\delta}+3\alpha+\bar{\beta}-\bar{\tau}+4\pi)(\delta+4\beta-\tau)-3\Psi_{2}-2\Phi_{11}\right]^{A}\Psi_{4}^{B}=\tilde{T}_{4},
\label{waveeq02}
\end{align}
where we have put the tilde to the source $T_{4}$ in order to emphasize that we are taking the gauge \eqref{gauge02} which is different from
\eqref{gauge01}, and then $T_{4}$ in \eqref{waveeq01} and that in \eqref{waveeq02} are different.

One can find that by taking different gauges, the last term in the left hand side of the wave equations can be changed,
namely $+2\Phi_{11}^{A}\Psi_{4}^{B}$ for \eqref{waveeq01} and $-2\Phi_{11}^{A}\Psi_{4}^{B}$ for \eqref{waveeq02}.
Here we will introduce yet another gauge such that the last term is $-6\Lambda^{A}\Psi_{4}^{B}$,
and we refer to as the \textit{separable gauge}.
One can rewrite \eqref{eq05} as
\begin{align}
&\left[(\Delta+3\gamma-\bar{\gamma}+4\mu+\bar{\mu})(D+4\epsilon-\rho)\right. \notag\\
&\qquad {}-\left.(\bar{\delta}+3\alpha+\bar{\beta}-\bar{\tau}+4\pi)(\delta+4\beta-\tau)-3\Psi_{2}-6\Lambda\right]^{A}\Psi_{4}^{B} \notag\\
&=T_{4} -4\lambda^{B}[(\Delta+\mu+\bar{\mu})\Phi_{11}]^{A}
-2(\Phi_{11}-3\Lambda)^{A}(\Delta+3\gamma-\bar{\gamma}+\mu+\bar{\mu})^{A}\lambda^{B} \notag\\
&\qquad {}-4\nu^{B}[(\bar{\delta}+\pi-\bar{\tau})\Phi_{11}]^{A}
-2(\Phi_{11}+3\Lambda)^{A}(\bar{\delta}+3\alpha+\bar{\beta}-\bar{\tau}+\pi)^{A}\nu^{B},
\label{eq08}
\end{align}
Then the gauge condition for the separable gauge is
\begin{align}
\Omega_{3}&\equiv\frac{1}{2}(\Phi_{11}-3\Lambda)^{A}(\Delta+3\gamma-\bar{\gamma}+\mu+\bar{\mu})^{A}\lambda^{B}
+\lambda^{B}[(\Delta+\mu+\bar{\mu})\Phi_{11}]^{A}
\notag\\
&\quad {}+\frac{1}{2}(\Phi_{11}+3\Lambda)^{A}(\bar{\delta}+3\alpha+\bar{\beta}-\bar{\tau}+\pi)^{A}\nu^{B}+\nu^{B}[(\bar{\delta}+\pi-\bar{\tau})\Phi_{11}]^{A}=0,
\label{gauge03}
\end{align}
which can also be rewritten as
\begin{gather}
\nu^{B}= -\frac{(\Delta+3\gamma-\bar{\gamma}+2\mu+2\bar{\mu})^{A}(\Phi_{11}^{A}\lambda^{B})}{(\bar{\delta}+\pi-\bar{\tau})^{A}\Phi_{11}^{A}}
-\frac{1}{2}\frac{(\Phi_{11}+3\Lambda)^{A}\Psi_{4}^{B}}{(\bar{\delta}+\pi-\bar{\tau})^{A}\Phi_{11}^{A}},
\label{gauge003-1}
\end{gather}
or
\begin{gather}
\lambda^{B}= -\frac{(\bar{\delta}+3\alpha+\bar{\beta}-2\bar{\tau}+2\pi)^{A}(\Phi_{11}^{A}\nu^{B})}{(\Delta+\mu+\bar{\mu})^{A}\Phi_{11}^{A}}
+\frac{1}{2}\frac{(\Phi_{11}-3\Lambda)^{A}\Psi_{4}^{B}}{(\Delta+\mu+\bar{\mu})^{A}\Phi_{11}^{A}}.
\label{gauge003-2}
\end{gather}
The corresponding wave equation is
\begin{align}
&\left[(\Delta+3\gamma-\bar{\gamma}+4\mu+\bar{\mu})(D+4\epsilon-\rho)\right. \notag\\
&\qquad {}-\left.(\bar{\delta}+3\alpha+\bar{\beta}-\bar{\tau}+4\pi)(\delta+4\beta-\tau)-3\Psi_{2}-6\Lambda\right]^{A}\Psi_{4}^{B}=\hat{T}_{4},
\label{waveeq03}
\end{align}
where we have put the hat to the source in a similar sense as \eqref{waveeq02}.
The advantage of the separable gauge is that the wave equation \eqref{waveeq03} %with no source $\hat{T}_{4}=0$
admits the separation of the variables. We define $\psi_{(-2)}$ and $T_{(-2)}$ as
\begin{gather}
\psi_{(-2)}=(\rho^{A})^{-4}\Psi_{4}^{B}=\bar{\varrho}^{4}\Psi_{4}^{B},\qquad T_{(-2)}=\bar{\varrho}^{4}\hat{T}_{4}.
\end{gather}
Then the wave equation for $\psi_{(-2)}$ is
\begin{align}
&\left[\frac{(r^{2}+a^{2})^{2}}{L(r)}-a^{2}\sin^{2}\theta\right]\frac{\partial^{2}\psi_{(-2)}}{\partial t^{2}}
+\left[\frac{2(r^{2}+a^{2})L'(r)}{L(r)}-8r-4ia\cos\theta\right]\frac{\partial\psi_{(-2)}}{\partial t}
\notag\\
&{}-L^{2}(r)\frac{\partial}{\partial r}\left(\frac{1}{L(r)}\frac{\partial\psi_{(-2)}}{\partial r}\right)
-\frac{1}{\sin\theta}\frac{\partial}{\partial\theta}\left(\sin\theta\frac{\partial\psi_{(-2)}}{\partial\theta}\right)
+\left(\frac{a^{2}}{L(r)}-\frac{1}{\sin^{2}\theta}\right)\frac{\partial^{2}\psi_{(-2)}}{\partial\varphi^{2}}
\notag\\
&{}+\left[\frac{2a(r^{2}+a^{2})}{L(r)}-2a\right]\frac{\partial^{2}\psi_{(-2)}}{\partial t \partial\varphi}
+\left(\frac{2aL'(r)}{L(r)}+\frac{4i\cos\theta}{\sin^2\theta}\right)\frac{\partial\psi_{(-2)}}{\partial\varphi}
\notag\\
&{}+(4\cot^{2}\theta+2)\psi_{(-2)}=2\Sigma T_{(-2)},
\label{PDE}
\end{align}
which reduces to the Teukolsky master equation on the Kerr background with the spin $s=-2$ in the case of $L=r^{2}-2Mr+a^{2}$.
\textcolor{black}{In the homogeneous case $T_{(-2)}=0$, }
for the separation of the variables, we assume the product form of the solution:
\begin{gather}
\psi_{(-2)}=e^{-i\omega t}e^{im\varphi}R(r)S(\theta),
\end{gather}
where $\omega$ is the frequency of the gravitational wave and $m$ is constant.
By substituting the above into \eqref{PDE}, we obtain the separated equations as
\begin{gather}
L^{2}\frac{d}{dr}\left(\frac{1}{L}\frac{dR}{dr}\right)
+\left(\frac{K^{2}+2iKL'}{L}-8i\omega r+2am\omega-a^{2}\omega^{2}-\boldsymbol{\lambda}_{(-2)}\right)R=0,
\label{defTE1}
\\
\begin{aligned}
&\frac{1}{\sin\theta}\frac{d}{d\theta}\left(\sin\theta\frac{dS}{d\theta}\right)
\\
&{}\qquad +\left(a^{2}\omega^{2}\cos^{2}\theta-\frac{m^{2}}{\sin^{2}\theta}+4a\omega\cos\theta+\frac{4m\cos\theta}{\sin^{2}\theta}
-4\cot^2\theta-2+\boldsymbol{\lambda}_{(-2)}\right)S=0,
\label{swsh}
\end{aligned}
\end{gather}
where $K=(r^{2}+a^{2})\omega -am$ and $\boldsymbol{\lambda}_{(-2)}$ is the separation constant.
From \eqref{swsh} one can find that $S(\theta)e^{im\varphi}$ coincides with the $s=-2$ spin-weighted spheroidal harmonics
${}^{}_{-2}S^{a\omega}_{lm}(\theta,\varphi)$, where $l$ and $m$ take the integer values with $l\ge 2$ and $-l\le m\le l$, respectively.
For the value of the separation constant $\boldsymbol{\lambda}_{(-2)}$, which becomes the eigenvalue of the spheroidal harmonics,
the exact expression is not known yet. Instead some approximated expressions are studied. For example, the small-$a\omega$ expansion
of $\boldsymbol{\lambda}_{(s)}$ for the general spin $s$ has been computed as \cite{Seidel:1988ue, Berti:2005gp}
%\footnote{Here we have shown up to the second order of $a\omega$ for simplicity, but some higher orders are also known. }
\begin{gather}
\boldsymbol{\lambda}_{(s)}=(l-s)(l+s+1)-\frac{2s^{2}m}{l(l+1)}a\omega+\left[h(l+1)-h(l)-1\right]a^{2}\omega^{2}+O(a^{3}\omega^{3}),
\label{eigenvalue}
\end{gather}
where $h(l)$ is given by
\begin{gather}
h(l)=\frac{2(l^{2}-m^{2})(l^{2}-s^{2})^{2}}{(2l-1)l^{3}(2l+1)}.
\end{gather}
The leading term $(l-s)(l+s+1)$ in \eqref{eigenvalue} is the eigenvalue of the spin-weighted spherical harmonics with the spin $s$.
\textcolor{black}{In the} nonhomogeneous case $T_{(-2)}\neq 0$, \textcolor{black}{it is useful to}
expand $\psi_{(-2)}$ and $T_{(-2)}$ by ${}^{}_{-2}S^{a\omega}_{lm}(\theta,\varphi)$ \textcolor{black}{which spans the complete basis} as
\begin{align}
\psi_{(-2)}&=\int d\omega \sum_{l,m} R^{(-2)}_{lm\omega}(r){}^{}_{-2}S^{a\omega}_{lm}(\theta,\varphi)e^{-i\omega t},
\\
-2\Sigma T_{(-2)}&=\int d\omega \sum_{l,m} G^{(-2)}_{lm\omega}(r){}^{}_{-2}S^{a\omega}_{lm}(\theta,\varphi)e^{-i\omega t}.
\end{align}
Then we obtain the ordinary differential equation for the radial coordinate as
\begin{gather}
L^{2}\frac{d}{dr}\left(\frac{1}{L}\frac{dR^{(-2)}_{lm\omega}}{dr}\right)
+\left(\frac{K^{2}+2iKL'}{L}-8i\omega r+2am\omega-a^{2}\omega^{2}-\boldsymbol{\lambda}_{(-2)}\right)R^{(-2)}_{lm\omega}=G^{(-2)}_{lm\omega},
\label{defTE2}
\end{gather}
which reduces to the Teukolsky radial equation on the Kerr background with the spin $s=-2$ in the case of $L=r^{2}-2Mr+a^{2}$,
as well as the master equation.

A similar analysis can be applied for the wave equation for $\Psi_{0}^{B}$, which in the separable gauge is obtained as
\begin{align}
&\left[(D-3\epsilon+\bar{\epsilon}-4\rho-\bar{\rho})(\Delta-4\gamma+\mu)\right.
\notag\\
&\qquad {}-\left.(\delta+\bar{\pi}-\bar{\alpha}-3\beta-4\tau)(\bar{\delta}+\pi-4\alpha)-3\Psi_{2}-6\Lambda\right]^{A}\Psi_{0}^{B}=T_{0},
\end{align}
where the source $T_{0}$ is defined as
\begin{align}
T_{0}&=(\delta+\bar{\pi}-\bar{\alpha}-3\beta-4\tau)^{A}
\left[(D-2\epsilon-2\bar{\rho})^{A}\Phi_{01}^{B}-(\delta+\bar{\pi}-2\bar{\alpha}-2\beta)^{A}\Phi_{00}^{B}\right]
\notag\\
&\quad {}+(D-3\epsilon+\bar{\epsilon}-4\rho-\bar{\rho})^{A}\left[(\delta+2\bar{\pi}-2\beta)^{A}\Phi_{01}^{B}
-(D-2\epsilon+2\bar{\epsilon}-\bar{\rho})^{A}\Phi_{02}^{B}\right].
\end{align}
The gauge condition is
\begin{align}
&\frac{1}{2}(\Phi_{11}-3\Lambda)^{A}(D-3\epsilon+\bar{\epsilon}-\rho-\bar{\rho})^{A}\sigma^{B}
+\sigma^{B}[(D-\rho-\bar{\rho})\Phi_{11}]^{A}
\notag\\
&\quad {}+\frac{1}{2}(\Phi_{11}+3\Lambda)^{A}(\delta+\bar{\pi}-\bar{\alpha}-3\beta-\tau)^{A}\kappa^{B}
+\kappa^{B}[(\delta+\bar{\pi}-\tau)\Phi_{11}]^{A}=0,
\end{align}
Then the explicit form of the wave equation  in turn becomes
\begin{align}
&\left[\frac{(r^{2}+a^{2})^{2}}{L(r)}-a^{2}\sin^{2}\theta\right]\frac{\partial^{2}\Psi_{0}^{B}}{\partial t^{2}}
+\left[-\frac{2(r^{2}+a^{2})L'(r)}{L(r)}+8r+4ia\cos\theta\right]\frac{\partial\Psi_{0}^{B}}{\partial t}
\notag\\
&{}-\frac{1}{L^{2}(r)}\frac{\partial}{\partial r}\left(L^{3}(r)\frac{\partial\Psi_{0}^{B}}{\partial r}\right)
-\frac{1}{\sin\theta}\frac{\partial}{\partial\theta}\left(\sin\theta\frac{\partial\Psi_{0}^{B}}{\partial\theta}\right)
+\left(\frac{a^{2}}{L(r)}-\frac{1}{\sin^{2}\theta}\right)\frac{\partial^{2}\Psi_{0}^{B}}{\partial\varphi^{2}}
\notag\\
&{}+\left[\frac{2a(r^{2}+a^{2})}{L(r)}-2a\right]\frac{\partial^{2}\Psi_{0}^{B}}{\partial t \partial\varphi}
-\left(\frac{2aL'(r)}{L(r)}+\frac{4i\cos\theta}{\sin^2\theta}\right)\frac{\partial\Psi_{0}^{B}}{\partial\varphi}
\notag\\
&{}+\left[4\cot^{2}\theta+2-2L''(r)\right]\Psi_{0}^{B}=2\Sigma T_{0},
\label{PDE2}
\end{align}
which reduces to the Teukolsky master equation with $s=2$ in the case of Kerr background.
After the separation of variables, the solution to the angular part gives the  $s=-2$ spin-weighted spheroidal harmonics
${}^{}_{2}S^{a\omega}_{lm}(\theta,\varphi)$.
On the other hand, the equation for the radial part becomes
\begin{align}
&\frac{1}{L^{2}}\frac{d}{dr}\left(L^{3}\frac{dR^{(2)}_{lm\omega}}{dr}\right)
\notag\\
&\qquad{}+\left(\frac{K^{2}-2iKL'}{L}+8i\omega r-4+2L''+2am\omega-a^{2}\omega^{2}-\boldsymbol{\lambda}_{(2)}\right)R^{(2)}_{lm\omega}
=G^{(2)}_{lm\omega},
\label{defTE2-2}
\end{align}
where we have used the following expansion
\begin{align}
\Psi_{0}^{B}&=\int d\omega \sum_{l,m} R^{(2)}_{lm\omega}(r){}^{}_{2}S^{a\omega}_{lm}(\theta,\varphi)e^{-i\omega t},
\\
-2\Sigma T_{0}&=\int d\omega \sum_{l,m} G^{(2)}_{lm\omega}(r){}^{}_{2}S^{a\omega}_{lm}(\theta,\varphi)e^{-i\omega t}.
\end{align}
We again note that \eqref{defTE2-2} reduces to the Teukolsky radial equation with $s=2$ in the case of Kerr background.

%\section{Teukolsky-like equations}

\section{Summary and discussion}\label{summary}

In this paper, we have studied a metric satisfying the axial symmetry, the \textcolor{black}{stationarity}, the Petrov type D property and the asymptotic flatness
with some additional ansatze, but without assuming the vacuum condition. The metric \eqref{metric7} in turn becomes the Kerr metric
deformed by the function $L(r)$ of the radial coordinate and the conformal factor $A(r,\theta)$.
We then have studied the gravitational-wave equation on the background of this metric in the case of $A(r,\theta)=1$.
For non-vacuum case, the wave equation is usually gauge-dependent, although the unknown variable $\Psi_{4}^{B}$ is gauge-invariant.
We have found one convenient gauge, refered to as the separable gauge, such that the \textcolor{black}{homogeneous} wave equation
admits the separation of the variables.
The solution of the equation \eqref{swsh} for the angular part just becomes the spin-weighted spheroidal harmonics.
The radial part \eqref{defTE2} of the equation gives the natural extension of the Teukolsky radial equation.
\textcolor{black}{This result is also helpful for solving the nonhomogeneous equation, because one can now recognize that it is appropriate to
expand the unknown function and the source with the spheroidal harmonics, and then the radial equation with the source can be obtained.  }

The metric \eqref{metric7} has been obtained using some assumptions, and hence it is not a classification.
In the case of \textcolor{black}{the} vacuum, the metric which posseses the type D property is completely classified,
called the Kinnersley metric \cite{Kinnersley:1969zza}.
It would be \textcolor{black}{interesting} to find the non-vacuum extension of the Kinnersley classification.
\textcolor{black}{
We also note that the metric \eqref{metric7} also contains the Kerr-Newman spacetime as a special case as $A(r,\theta)=1$,
$L(r)=r^{2}-2Mr+a^{2}+q^{2}$, where $q$ is the electric charge. In this case, we have $\Phi_{11}^{A}=q^{2}/(2\Sigma^{2})$, $\Lambda=0$.
Here nonvanishing $\Phi_{11}^{A}$ appears due to the contribution from the background electromagnetic field. Unfortunately, our analysis may not
help to solve the long-standing problem about the (non-)separability in the wave equation, because we do not consider the electromagnetic perturbation, and it has been known that the problem happens when the gravitational and the electromagnetic perturbations are considered simultaneously.
On the other hand, it has also been known that there is no such problem in the Reissner-Nordstr\"om spacetime which is spherically symmetric.
It would be interesting to consider the case with the coexistence of the gravitational and the electromagnetic perturbation in the background
of the non-vacuum spherically symmetric spacetime.
}

We have considered the gravitational-wave equation only for the special case $A(r,\theta)=1$ of the background, because otherwise
all of the components of the Ricci tensor ($\Phi^{A}$'s and $\Lambda^{A}$) becomes nonzero, and then it is complicated.
We will leave to study this issue as our future work.
The wave equations \eqref{PDE} and \eqref{defTE2} have been obtained under the separable gauge \eqref{gauge03}.
Since the gauge dependence of the wave equation is due to that of the source term $T_{4}$, in particular $\Phi_{21}^{B}$ as \eqref{source} below,
the gauge choice seems to be unavoidable. Then we have to consider how to ensure the gauge condition, which would actually be the problem
for any type of the gauge condition.

The property of the wave equations derived here %\eqref{PDE} and \eqref{defTE2}
strongly depends on the behaviour of the function $L(r)$.
In particular, the zeroes of $L(r)$ would give some of the singularites of the equation \eqref{defTE2}.
For example, in the case of the Kerr background $L(r)=r^{2}-2Mr+a^{2}$, the zeroes of $L(r)$ are just located at the inner and the outer horizon,
which appear as the two regular singularities of the Teukolsky equation. Since the Teukolsky equation also has one irregular singularity
at the infinity $r\to\infty$, it is mathematically classified as an example of the confluent Heun equation \cite{Heun}.
Similarly, in the present case, the behaviour of $L(r)$ is very important. For the large-$r$ behavior, since $L(r)\sim r^{2}$
from the asymptotic flatness, the structure of the wave equation is similar with that of the Teukolsky equation.
However, for the small $r$, $L(r)$ could have the behaviour different from the Kerr case. In particular, the number of the zeroes
could change, which gives the quite different behaviour of the wave equation, because the zeroes of $L(r)$ could give the singularities
of the differential equation.
%\footnote{In the Kerr case, two zeroes of $L(r)=r^{2}-2Mr+a^{2}$ gives the regular singularities of the Teukolsky equation
%which also has one irregular singularity at $r\to\infty$. Then the Teukolsky equation is classified as an example of
%the confluent Heun equation. }.
The small-$r$ behaviour is very important for the scattering problem as well,
since the boundary condition of the wave is determined from that.
It is also important to calculate the quasi-normal frequency and the fluxes of the gravitational waves
\cite{Cardoso:2021wlq, Cardoso:2022whc, Destounis:2022obl, Figueiredo:2023gas}.

Finally here we have considered only the gravitational-wave equation. It would be interesting to study the wave equation on this background with
the different spin, namely the Klein-Gordon equation, the Dirac equation and the Maxwell equation,
as in the case of the spherically symmetric background \cite{Guo:2023hdn}.

\section*{Acknowledgements}
This work was supported in part by the National Natural Science Foundation of China (Grant No. 11973025).

\begin{appendix}

%\section{Goldberg-Sachs theorem and its non-vacuum extension}\label{GS}
%
%In Section 2, first we have required $\kappa=\sigma=\lambda=\nu=0$\footnote{Here the quantities are all in the background,
%and then the superscript $A$ will be omitted for notational simplicity.}, and then have found the solution to
%the type D condition $\Psi_{0}=\Psi_{1}=\Psi_{3}=\Psi_{4}=0$. In the vacuum background, the former implies the latter,
%which is called the Goldberg-Sachs theorem.

\section{Reduction of the background to spherically symmetric case}\label{sphesym}

In the case of $A(r,\theta)=A(r)$ and $a=0$, the metric \eqref{metric7} takes the following form:
\begin{gather}
ds^{2}=A(r)\left[\frac{L(r)}{r^{2}}dt^{2}-\frac{r^{2}}{L(r)}dr^{2}-r^{2}(d\theta^{2}+\sin^{2}\theta d\varphi^{2})\right].
\label{metric8}
\end{gather}
On the other hand, the general form of the metric for the spherically symmetric background is
\begin{gather}
ds^{2}=\mathcal{A}(r)dt^{2}-\mathcal{B}(r)dr^{2}-\mathcal{C}(r)r^{2}(d\theta^{2}+\sin^{2}\theta d\varphi^{2}).
\end{gather}
From the comparison of two metrics, we have
\begin{gather}
\mathcal{A}(r)=\frac{A(r)L(r)}{r^{2}}, \quad \mathcal{B}(r)=\frac{r^{2}A(r)}{L(r)}, \quad \mathcal{C}(r)=A(r).
\end{gather}
By eliminating $A(r)$ and $L(r)$ from the above, we obtain
\begin{gather}
\mathcal{C}(r)=\sqrt{\mathcal{A}(r)\mathcal{B}(r)},
\end{gather}
which can be regarded as the gauge choice of radial coordinate, and the metric \eqref{metric7} contains the general case of
the spherically symmetric background. In the previous study, we have used the different coordinate such that $\mathcal{C}(r)$=1.
In the case of $A(r)=1$, which gives $\mathcal{A}(r)\mathcal{B}(r)=1$, these two choices agree.

\section{Gauge condition}\label{gaugecon}

In this paper, we have studied a few types of the gauge condition, which is imposed in order to obtain the decoupled equation of the gravitational wave.
At the beginning, we have four variables $\Psi_{4}^{B}$, $\Psi_{3}^{B}$, $\lambda^{B}$ and $\nu^{B}$ in \eqref{eq01}--\eqref{eq03}.
Among them $\Psi_{3}^{B}$ is eliminated by using the commutation relation \eqref{com}, and then the three variables remain.
These variables are not independent and there is one relation \eqref{eq03} between them, which is gauge-covariant.
Hence we can have another one relation as the gauge conditon to obtain the decoupled equation for one variable.
Here we will use the tetrad rotation (the local Lorentz transformation) as the gauge symmetry,
which consists of the following three kinds \cite{Janis:1965tx}:
\begin{align}
& l^{\mu}\to l^{\mu},\quad m^{\mu}\to m^{\mu}+al^{\mu}, \quad \bar{m}^{\mu}\to \bar{m}^{\mu}+\bar{a}l^{\mu},
\quad n^{\mu}\to n^{\mu}+\bar{a}m^{\mu}+a\bar{m}^{\mu}+a\bar{a}l^{\mu},
\label{rot01}\\
& n^{\mu}\to n^{\mu},\quad m^{\mu}\to m^{\mu}+bn^{\mu}, \quad \bar{m}^{\mu}\to \bar{m}^{\mu}+\bar{b}n^{\mu},
\quad l^{\mu}\to l^{\mu}+\bar{b}m^{\mu}+b\bar{m}^{\mu}+b\bar{b}n^{\mu},
\label{rot02}\\
& l^{\mu}\to e^{-c}l^{\mu},\quad n^{\mu}\to e^{c}n^{\mu},\quad
m^{\mu}\to e^{i\vartheta}m^{\mu},\quad \bar{m}^{\mu}\to e^{-i\vartheta}\bar{m}^{\mu},
\label{rot03}
\end{align}
where  $a$ and $b$ are the complex functions\footnote{Only here $a$ is not the Kerr parameter. We hope the readers may not confuse it. }
and $c$ and $\vartheta$ are real functions.
Since we do not want to change the background, these parameter functions has to be regarded as the perturbation quantities, and
the linear order of the transformation is enough.
The variables $\lambda^{B}$ and $\nu^{B}$ are transformed as\footnote{$\lambda^{B}$ and $\nu^{B}$ are invariant
under \eqref{rot02} and \eqref{rot03}. A similar thing also holds for \eqref{source}. }
\begin{gather}
\lambda^{B}\to\lambda^{B}+(\bar{\delta}+2\alpha+\pi)^{A}\bar{a}, \quad \nu^{B}\to\nu^{B}+(\Delta+2\gamma+\mu)^{A}\bar{a},
\end{gather}
whereas $\Psi_{4}^{B}$ is invariant, which can also be checked using the commutation relation \eqref{com} with $p=2$, $q=-1$.

The source term $T_{4}$ contains the perturbation quantities $\Phi_{20}^{B}$, $\Phi_{21}^{B}$ and $\Phi_{22}^{B}$, which
transforms under \eqref{rot01}--\eqref{rot03} as
\begin{gather}
\Phi_{20}^{B}\to\Phi_{20}^{B}, \quad \Phi_{21}^{B}\to\Phi_{21}^{B}+2\Phi_{11}^{A}\bar{a}, \quad \Phi_{22}^{B}\to\Phi_{22}^{B}.
\label{source}
\end{gather}
From the above, $T_{4}$ transforms as
\begin{align}
T_{4}&\to T_{4}+4(\bar{\delta}+3\alpha+\bar{\beta}-\bar{\tau}+4\pi)^{A}(\Delta+2\gamma+2\mu+\bar{\mu})^{A}(\Phi_{11}^{A}\bar{a})
\notag\\
&\qquad\ \ {}-4(\Delta+3\gamma-\bar{\gamma}+4\mu+\bar{\mu})^{A}[(\bar{\tau}+2\pi)^{A}\Phi_{11}^{A}\bar{a}].
\end{align}
The gauge dependence of the wave equations is due to the gauge dependence of $T_{4}$, although $T_{4}$ is written in terms of the
energy-momentum tensor of the matter through the Einstein equation.
The above transformation has to be canceled by that of terms in the right hand side in \eqref{waveeq01}, \eqref{waveeq02} and \eqref{waveeq03}
other than $T_{4}$, because the left hand side of each equation is gauge-invariant. Then the gauge transformation of the gauge conditions
\eqref{gauge01}, \eqref{gauge02} and \eqref{gauge03} is
\begin{align}
\Omega_{i}&\to\Omega_{i}+(\bar{\delta}+3\alpha+\bar{\beta}-\bar{\tau}+4\pi)^{A}(\Delta+2\gamma+2\mu+\bar{\mu})^{A}(\Phi_{11}^{A}\bar{a})
\notag\\
&\qquad\ \ {}-(\Delta+3\gamma-\bar{\gamma}+4\mu+\bar{\mu})^{A}[(\bar{\tau}+2\pi)^{A}\Phi_{11}^{A}\bar{a}],
\label{transf}
\end{align}
where $i$ runs from 1 to 3, corresponding to the three gauge conditions.
The gauge parameter $a$ has to be chosen so that the right hand side of \eqref{transf} vanishes.

\end{appendix}

\section*{Acknowledgements}
The authors thank Prof. Ruffini for useful discussions. This work was supported in part by the National Natural Science Foundation of China (Grant No. 11973025).


\begin{thebibliography}{99}

%\cite{Abbott:2016blz}
\bibitem{Abbott:2016blz}
B.~Abbott \textit{et al.} [LIGO Scientific and Virgo],
%``Observation of Gravitational Waves from a Binary Black Hole Merger,''
Phys. Rev. Lett. \textbf{116}, no.6, 061102 (2016)
doi:10.1103/PhysRevLett.116.061102
[arXiv:1602.03837 [gr-qc]].
%5183 citations counted in INSPIRE as of 22 Jun 2020

%\cite{PoissonWill}
\bibitem{PoissonWill}
E.~Poisson and C.~M.~Will,
``Gravity: Newtonian, Post-Newtonian, Relativistic.''
Cambridge University Press (2014)
doi:10.1017/CBO9781139507486.

%\cite{Mino:1997bx}
\bibitem{Mino:1997bx}
Y.~Mino, M.~Sasaki, M.~Shibata, H.~Tagoshi and T.~Tanaka,
%``Black hole perturbation: Chapter 1,''
Prog. Theor. Phys. Suppl. \textbf{128}, 1-121 (1997)
doi:10.1143/PTPS.128.1
[arXiv:gr-qc/9712057 [gr-qc]].
%107 citations counted in INSPIRE as of 22 Jun 2020

%\cite{Teukolsky:1973ha}
\bibitem{Teukolsky:1973ha}
  S.~A.~Teukolsky,
  %``Perturbations of a rotating black hole. 1. Fundamental equations for gravitational electromagnetic and neutrino field perturbations,''
  Astrophys.\ J.\  {\bf 185}, 635 (1973)
  doi:10.1086/152444.
  %%CITATION = doi:10.1086/152444;%%
  %1061 citations counted in INSPIRE as of 27 Jul 2019

%\cite{Buonanno:2000ef}
\bibitem{Buonanno:2000ef}
A.~Buonanno and T.~Damour,
%``Transition from inspiral to plunge in binary black hole coalescences,''
Phys. Rev. D \textbf{62}, 064015 (2000)
doi:10.1103/PhysRevD.62.064015
[arXiv:gr-qc/0001013 [gr-qc]].
%641 citations counted in INSPIRE as of 19 Dec 2022

%\cite{Damour:2016gwp}
\bibitem{Damour:2016gwp}
T.~Damour,
%``Gravitational scattering, post-Minkowskian approximation and Effective One-Body theory,''
Phys. Rev. D \textbf{94}, no.10, 104015 (2016)
doi:10.1103/PhysRevD.94.104015
[arXiv:1609.00354 [gr-qc]].
%190 citations counted in INSPIRE as of 19 Dec 2022

%\cite{Damour:2001tu}
\bibitem{Damour:2001tu}
T.~Damour,
%``Coalescence of two spinning black holes: an effective one-body approach,''
Phys. Rev. D \textbf{64}, 124013 (2001)
doi:10.1103/PhysRevD.64.124013
[arXiv:gr-qc/0103018 [gr-qc]].
%501 citations counted in INSPIRE as of 18 Aug 2023

%\cite{Petrov}
\bibitem{Petrov}
A.~Z.~Petrov,
``Classification of spaces defined by gravitational fields,''
Uch. Zapiski Kazan Gos. Univ. 144, 55 (1954).

%\cite{Jing:2021ahx}
\bibitem{Jing:2021ahx}
J.~Jing, S.~Chen, M.~Sun, X.~He, M.~Wang and J.~Wang,
%``Self-consistent effective-one-body theory for spinless binaries based on post-Minkowskian approximation I: Hamiltonian and decoupled equation for $\psi _4^{\rm{B}}$,''
Sci. China Phys. Mech. Astron. \textbf{65}, no.6, 260411 (2022)
doi:10.1007/s11433-022-1885-6
[arXiv:2112.09838 [gr-qc]].
%2 citations counted in INSPIRE as of 19 Dec 2022

%\cite{Jing:2022vks}
\bibitem{Jing:2022vks}
J.~Jing, S.~Long, W.~Deng, M.~Wang and J.~Wang,
%``New self-consistent effective one-body theory for spinless binaries based on the post-Minkowskian approximation,''
Sci. China Phys. Mech. Astron. \textbf{65}, no.10, 100411 (2022)
doi:10.1007/s11433-022-1951-1
[arXiv:2208.02420 [gr-qc]].
%0 citations counted in INSPIRE as of 19 Dec 2022

%\cite{Guo:2023niy}
\bibitem{Guo:2023niy}
Y.~Guo, H.~Nakajima and W.~Lin,
%``Gravitational-wave equation in effective one-body background for spinless binary,''
Sci. China Phys. Mech. Astron. \textbf{66}, no.7, 270412 (2023)
doi:10.1007/s11433-023-2087-8
[arXiv:2301.08318 [gr-qc]].
%0 citations counted in INSPIRE as of 15 Aug 2023
%
%%\cite{Guo2023-2}
%\bibitem{Guo2023-2}
%Y.~Guo, H.~Nakajima and W.~Lin,
%``Teukolsky-like equations with various spins in spherically symmetric spacetime.''[arXiv:2309.04758 [gr-qc]].

%\cite{Guo:2023hdn}
\bibitem{Guo:2023hdn}
Y.~Guo, H.~Nakajima and W.~Lin,
``Teukolsky-like equations with various spins in spherically symmetric spacetime,''
[arXiv:2309.04758 [gr-qc]].
%1 citations counted in INSPIRE as of 18 Sep 2023

%\cite{Jing:2023vzq}
\bibitem{Jing:2023vzq}
J.~Jing, W.~Deng, S.~Long and J.~Wang,
%``Self-consistent effective-one-body theory for spinning binaries based on post-Minkowskian approximation,''
Sci. China Phys. Mech. Astron. \textbf{66}, no.7, 270411 (2023)
doi:10.1007/s11433-023-2084-1
[arXiv:2305.03225 [gr-qc]].
%4 citations counted in INSPIRE as of 18 Aug 2023

%\cite{Newman:1961qr}
\bibitem{Newman:1961qr}
  E.~Newman and R.~Penrose,
  %``An Approach to gravitational radiation by a method of spin coefficients,''
  J.\ Math.\ Phys.\  {\bf 3}, 566 (1962)
  doi:10.1063/1.1724257.
%  %%CITATION = doi:10.1063/1.1724257;%%
%  %1192 citations counted in INSPIRE as of 27 Jul 2019

%\cite{GS}
\bibitem{GS}
J.~N.~Goldberg and R.~K.~Sachs,
%``A theorem on Petrov types,''
Acta Phys. Polon. Suppl. \textbf{22}, 13 (1962).

%\cite{KT}
\bibitem{KT}
W.~Kundt and A.~Thompson,
%``Le tenseur de Weyl et une congruence associ\'{e}e de g\'{e}od\'{e}siques isotropes sans distorsion,''
C. R. Acad. Sci. (Paris) \textbf{254}, 4257 (1962).

%\cite{RS}
\bibitem{RS}
I.~Robinson and A.~Schild,
%``Generalization of a theorem by Goldberg and Sachs,''
J. Math. Phys. \textbf{4}, 484 (1963).

%\cite{Stephani:2003tm}
\bibitem{Stephani:2003tm}
H.~Stephani, D.~Kramer, M.~A.~H.~MacCallum, C.~Hoenselaers and E.~Herlt,
%``Exact solutions of Einstein's field equations,''
Cambridge Univ. Press, 2003,
ISBN 978-0-521-46702-5, 978-0-511-05917-9
doi:10.1017/CBO9780511535185
%574 citations counted in INSPIRE as of 21 Aug 2023

%\cite{Seidel:1988ue}
\bibitem{Seidel:1988ue}
E.~Seidel,
%``A Comment on the Eigenvalues of Spin Weighted Spheroidal Functions,''
Class. Quant. Grav. \textbf{6}, 1057 (1989)
doi:10.1088/0264-9381/6/7/012
%72 citations counted in INSPIRE as of 18 Sep 2023

%\cite{Berti:2005gp}
\bibitem{Berti:2005gp}
E.~Berti, V.~Cardoso and M.~Casals,
%``Eigenvalues and eigenfunctions of spin-weighted spheroidal harmonics in four and higher dimensions,''
Phys. Rev. D \textbf{73}, 024013 (2006)
[erratum: Phys. Rev. D \textbf{73}, 109902 (2006)]
doi:10.1103/PhysRevD.73.109902
[arXiv:gr-qc/0511111 [gr-qc]].
%312 citations counted in INSPIRE as of 18 Sep 2023

%\cite{Kinnersley:1969zza}
\bibitem{Kinnersley:1969zza}
W.~Kinnersley,
%``Type D Vacuum Metrics,''
J. Math. Phys. \textbf{10}, 1195-1203 (1969)
doi:10.1063/1.1664958
%321 citations counted in INSPIRE as of 17 Aug 2023

%\cite{Heun}
\bibitem{Heun}
A.~Ronveaux,
``Heun's Differential Equations,''
Oxford University Press, Oxford, New York, October 1995.

%%\cite{Chandrasekhar:1975}
%\bibitem{Chandrasekhar:1975}
%  S.~Chandrasekhar,
%  %``On the equations governing the perturbations of the Schwarzschild black hole,''
%   Proc. R. Soc. Lond. A \textbf{A343}, 289-298 (1975)
%   doi:10.1098/rspa.1975.0066
%
%\bibitem{Starobinsky1}
%A. A. Starobinsky and S. M. Churilov, Zh. Eksp. Teor. Fiz. \textbf{65}, 3,
%A. A. Starobinsky and S. M. Churilov, Sov. Phys. JETP 38, 1 (1974).

%\cite{Cardoso:2021wlq}
\bibitem{Cardoso:2021wlq}
V.~Cardoso, K.~Destounis, F.~Duque, R.~P.~Macedo and A.~Maselli,
%``Black holes in galaxies: Environmental impact on gravitational-wave generation and propagation,''
Phys. Rev. D \textbf{105}, no.6, L061501 (2022)
doi:10.1103/PhysRevD.105.L061501
[arXiv:2109.00005 [gr-qc]].
%60 citations counted in INSPIRE as of 18 Sep 2023

%\cite{Cardoso:2022whc}
\bibitem{Cardoso:2022whc}
V.~Cardoso, K.~Destounis, F.~Duque, R.~Panosso Macedo and A.~Maselli,
%``Gravitational Waves from Extreme-Mass-Ratio Systems in Astrophysical Environments,''
Phys. Rev. Lett. \textbf{129}, no.24, 241103 (2022)
doi:10.1103/PhysRevLett.129.241103
[arXiv:2210.01133 [gr-qc]].
%35 citations counted in INSPIRE as of 18 Sep 2023

%\cite{Destounis:2022obl}
\bibitem{Destounis:2022obl}
K.~Destounis, A.~Kulathingal, K.~D.~Kokkotas and G.~O.~Papadopoulos,
%``Gravitational-wave imprints of compact and galactic-scale environments in extreme-mass-ratio binaries,''
Phys. Rev. D \textbf{107}, no.8, 084027 (2023)
doi:10.1103/PhysRevD.107.084027
[arXiv:2210.09357 [gr-qc]].
%17 citations counted in INSPIRE as of 18 Sep 2023

%\cite{Figueiredo:2023gas}
\bibitem{Figueiredo:2023gas}
E.~Figueiredo, A.~Maselli and V.~Cardoso,
%``Black holes surrounded by generic dark matter profiles: Appearance and gravitational-wave emission,''
Phys. Rev. D \textbf{107}, no.10, 104033 (2023)
doi:10.1103/PhysRevD.107.104033
[arXiv:2303.08183 [gr-qc]].
%7 citations counted in INSPIRE as of 18 Sep 2023

%\cite{Janis:1965tx}
\bibitem{Janis:1965tx}
A.~I.~Janis and E.~T.~Newman,
%``Structure of Gravitational Sources,''
J. Math. Phys. \textbf{6}, 902-914 (1965)
doi:10.1063/1.1704349
%80 citations counted in INSPIRE as of 21 Dec 2022






%%\cite{Chandrasekhar:1985kt}
%\bibitem{Chandrasekhar:1985kt}
%S.~Chandrasekhar,
%``The mathematical theory of black holes,''
%Springer (1985)
%doi:10.1007/978-94-009-6469-3\_2
%%341 citations counted in INSPIRE as of 19 Sep 2020

%%\cite{Thompson:2016fxe}
%\bibitem{Thompson:2016fxe}
%J.~E.~Thompson, B.~F.~Whiting and H.~Chen,
%%``Gauge Invariant Perturbations of the Schwarzschild Spacetime,''
%Class. Quant. Grav. \textbf{34}, no.17, 174001 (2017)
%doi:10.1088/1361-6382/aa7f5b
%[arXiv:1611.06214 [gr-qc]].
%%23 citations counted in INSPIRE as of 19 Dec 2022

%%\cite{Regge:1957td}
%\bibitem{Regge:1957td}
%  T.~Regge and J.~A.~Wheeler,
%  %``Stability of a Schwarzschild singularity,''
%  Phys.\ Rev.\  {\bf 108}, 1063 (1957)
%  doi:10.1103/PhysRev.108.1063.
%  %%CITATION = doi:10.1103/PhysRev.108.1063;%%
%  %1367 citations counted in INSPIRE as of 27 Jul 2019
%
%%\cite{Zerilli:1970se}
%\bibitem{Zerilli:1970se}
%  F.~J.~Zerilli,
%  %``Effective potential for even parity Regge-Wheeler gravitational perturbation equations,''
%  Phys.\ Rev.\ Lett.\  {\bf 24}, 737 (1970)
%  doi:10.1103/PhysRevLett.24.73.7.
%  %%CITATION = doi:10.1103/PhysRevLett.24.737;%%
%  %573 citations counted in INSPIRE as of 27 Jul 2019
%
%%\cite{Moncrief:1974am}
%\bibitem{Moncrief:1974am}
%  V.~Moncrief,
%  %``Gravitational perturbations of spherically symmetric systems. I. The exterior problem.,''
%  Annals Phys.\  {\bf 88}, 323 (1974)
%  doi:10.1016/0003-4916(74)90173-0.
%  %%CITATION = doi:10.1016/0003-4916(74)90173-0;%%
%  %302 citations counted in INSPIRE as of 27 Jul 2019

%%\cite{Detweiler:2008ft}
%\bibitem{Detweiler:2008ft}
%S.~L.~Detweiler,
%%``A Consequence of the gravitational self-force for circular orbits of the Schwarzschild geometry,''
%Phys. Rev. D \textbf{77}, 124026 (2008)
%doi:10.1103/PhysRevD.77.124026
%[arXiv:0804.3529 [gr-qc]].
%%202 citations counted in INSPIRE as of 19 Dec 2022


%%\cite{Baumgarte:2010ndz}
%\bibitem{Baumgarte:2010ndz}
%T.~W.~Baumgarte and S.~L.~Shapiro,
%``Numerical Relativity: Solving Einstein's Equations on the Computer,''
%Cambridge University Press (2010)
%doi:10.1017/CBO9781139193344
%%57 citations counted in INSPIRE as of 19 Sep 2020

%%\cite{Bardeen:1973xb}
%\bibitem{Bardeen:1973xb}
%  J.~Bardeen and W.~Press,
%  %``Radiation fields in the schwarzschild background,''
%  J. Math. Phys. \textbf{14}, 7-19 (1973)
%  doi:10.1063/1.1666175.
%  %153 citations counted in INSPIRE as of 22 Jun 2020

%%\cite{Lenzi:2021wpc}
%\bibitem{Lenzi:2021wpc}
%M.~Lenzi and C.~F.~Sopuerta,
%%``Master functions and equations for perturbations of vacuum spherically symmetric spacetimes,''
%Phys. Rev. D \textbf{104}, no.8, 084053 (2021)
%doi:10.1103/PhysRevD.104.084053
%[arXiv:2108.08668 [gr-qc]].
%%2 citations counted in INSPIRE as of 04 Dec 2021
%
%%\cite{Liu:2022csl}
%\bibitem{Liu:2022csl}
%W.~Liu, X.~Fang, J.~Jing and A.~Wang,
%%``Gauge invariant perturbations of general spherically symmetric spacetimes,''
%Sci. China Phys. Mech. Astron. \textbf{66}, no.1, 210411 (2023)
%doi:10.1007/s11433-022-1956-4
%[arXiv:2201.01259 [gr-qc]].
%%3 citations counted in INSPIRE as of 19 Dec 2022


%%\cite{Fujita:2014eta}
%\bibitem{Fujita:2014eta}
%R.~Fujita,
%%``Gravitational Waves from a Particle in Circular Orbits around a Rotating Black Hole to the 11th Post-Newtonian Order,''
%PTEP \textbf{2015}, no.3, 033E01 (2015)
%doi:10.1093/ptep/ptv012
%[arXiv:1412.5689 [gr-qc]].
%%21 citations counted in INSPIRE as of 27 Jun 2020


%%\cite{Detweiler:1977gy}
%\bibitem{Detweiler:1977gy}
%S.~L.~Detweiler,
%%``Resonant oscillations of a rapidly rotating black hole,''
%Proc. Roy. Soc. Lond. A \textbf{A352}, 381-395 (1977)
%doi:10.1098/rspa.1977.0005
%%56 citations counted in INSPIRE as of 26 Jun 2020
%
%%\cite{Chandrasekhar:1976zz}
%\bibitem{Chandrasekhar:1976zz}
%S.~Chandrasekhar and S.~L.~Detweiler,
%%``Equations governing gravitational perturbations of the Kerr black-hole,''
%Proc. Roy. Soc. Lond. A \textbf{A350}, 165-174 (1976)
%doi:10.1098/rspa.1976.0101.
%%35 citations counted in INSPIRE as of 22 Jun 2020
%
%%\cite{Sasaki:1981sx}
%\bibitem{Sasaki:1981sx}
%M.~Sasaki and T.~Nakamura,
%%``Gravitational Radiation From a Kerr Black Hole. 1. Formulation and a Method for Numerical Analysis,''
%Prog. Theor. Phys. \textbf{67}, 1788 (1982)
%doi:10.1143/PTP.67.1788
%%115 citations counted in INSPIRE as of 22 Jun 2020
%
%%\cite{Darboux}
%\bibitem{Darboux}
%  G.~Darboux,
%  C.R.Academy Sci.(Paris) 94 (1882) 1456
%  [arXiv:physics/9908003 [physics]].
%
%%\cite{Glampedakis:2017rar}
%\bibitem{Glampedakis:2017rar}
%K.~Glampedakis, A.~D.~Johnson and D.~Kennefick,
%%``Darboux transformation in black hole perturbation theory,''
%Phys. Rev. D \textbf{96}, no.2, 024036 (2017)
%doi:10.1103/PhysRevD.96.024036
%[arXiv:1702.06459 [gr-qc]].
%%7 citations counted in INSPIRE as of 22 Jun 2020
%
%%\cite{Yurov:2018ynn}
%\bibitem{Yurov:2018ynn}
%A.~V.~Yurov and V.~A.~Yurov,
%%``A look at the generalized Darboux transformations for the quasinormal spectra in Schwarzschild black hole perturbation theory: just how general should it be?,''
%Phys. Lett. A \textbf{383}, no.22, 2571-2578 (2019)
%doi:10.1016/j.physleta.2019.05.024
%[arXiv:1809.10279 [gr-qc]].
%%0 citations counted in INSPIRE as of 22 Jun 2020
%
%%\cite{Heun}
%\bibitem{Heun}
%A.~Ronveaux,
%``Heun's Differential Equations,''
%Oxford University Press, Oxford, New York, October 1995.
%
%%\cite{Mano:1996vt}
%\bibitem{Mano:1996vt}
%S.~Mano, H.~Suzuki and E.~Takasugi,
%%``Analytic solutions of the Teukolsky equation and their low frequency expansions,''
%Prog. Theor. Phys. \textbf{95}, 1079-1096 (1996)
%doi:10.1143/PTP.95.1079
%[arXiv:gr-qc/9603020 [gr-qc]].
%%155 citations counted in INSPIRE as of 19 Sep 2020
%
%%\cite{Casals:2021ugr}
%\bibitem{Casals:2021ugr}
%M.~Casals and R.~T.~da Costa,
%%``Hidden spectral symmetries and mode stability of subextremal Kerr(-dS) black holes,''
%arXiv:2105.13329 [gr-qc].
%%0 citations counted in INSPIRE as of 23 Oct 2021
%
%%\cite{Aminov:2020yma}
%\bibitem{Aminov:2020yma}
%G.~Aminov, A.~Grassi and Y.~Hatsuda,
%%``Black Hole Quasinormal Modes and Seiberg-Witten Theory,''
%arXiv:2006.06111 [hep-th].
%%18 citations counted in INSPIRE as of 22 Oct 2021
%
%%\cite{Hatsuda:2020iql}
%\bibitem{Hatsuda:2020iql}
%Y.~Hatsuda,
%%``An alternative to the Teukolsky equation,''
%arXiv:2007.07906 [gr-qc].
%%2 citations counted in INSPIRE as of 22 Oct 2021
%
%%\cite{Seiberg:1994rs}
%\bibitem{Seiberg:1994rs}
%N.~Seiberg and E.~Witten,
%%``Electric - magnetic duality, monopole condensation, and confinement in N=2 supersymmetric Yang-Mills theory,''
%Nucl. Phys. B \textbf{426}, 19-52 (1994)
%[erratum: Nucl. Phys. B \textbf{430}, 485-486 (1994)]
%doi:10.1016/0550-3213(94)90124-4
%[arXiv:hep-th/9407087 [hep-th]].
%%3426 citations counted in INSPIRE as of 25 Oct 2021
%
%%\cite{Seiberg:1994aj}
%\bibitem{Seiberg:1994aj}
%N.~Seiberg and E.~Witten,
%%``Monopoles, duality and chiral symmetry breaking in N=2 supersymmetric QCD,''
%Nucl. Phys. B \textbf{431}, 484-550 (1994)
%doi:10.1016/0550-3213(94)90214-3
%[arXiv:hep-th/9408099 [hep-th]].
%%2449 citations counted in INSPIRE as of 25 Oct 2021
%
%%\cite{NS}
%\bibitem{NS}
%N.~A.~Nekrasov and S.~L.~Shatashvili,
%``Quantization of integrable systems and four dimensional gauge theories,''
%XVIth International Congress on Mathematical Physics,
%World Scientific, March 2010, pp. 265-289.
%
%%\cite{Alday:2009aq}
%\bibitem{Alday:2009aq}
%L.~F.~Alday, D.~Gaiotto and Y.~Tachikawa,
%%``Liouville Correlation Functions from Four-dimensional Gauge Theories,''
%Lett. Math. Phys. \textbf{91}, 167-197 (2010)
%doi:10.1007/s11005-010-0369-5
%[arXiv:0906.3219 [hep-th]].
%%1150 citations counted in INSPIRE as of 25 Oct 2021
%
%%\cite{Gaiotto:2009ma}
%\bibitem{Gaiotto:2009ma}
%D.~Gaiotto,
%%``Asymptotically free $\mathcal{N} = 2$ theories and irregular conformal blocks,''
%J. Phys. Conf. Ser. \textbf{462}, no.1, 012014 (2013)
%doi:10.1088/1742-6596/462/1/012014
%[arXiv:0908.0307 [hep-th]].
%%230 citations counted in INSPIRE as of 25 Oct 2021
%
%%\cite{Bonelli:2021uvf}
%\bibitem{Bonelli:2021uvf}
%G.~Bonelli, C.~Iossa, D.~P.~Lichtig and A.~Tanzini,
%%``Exact solution of Kerr black hole perturbations via CFT$_2$ and instanton counting. Greybody factor,
%%Quasinormal modes and Love numbers,''
%[arXiv:2105.04483 [hep-th]].
%%7 citations counted in INSPIRE as of 23 Oct 2021
%
%%\cite{Whiting:1988vc}
%\bibitem{Whiting:1988vc}
%B.~F.~Whiting,
%%``Mode Stability of the Kerr Black Hole,''
%J. Math. Phys. \textbf{30}, 1301 (1989)
%doi:10.1063/1.528308
%%221 citations counted in INSPIRE as of 22 Oct 2021
%
%%\cite{Andersson:2016epf}
%\bibitem{Andersson:2016epf}
%L.~Andersson, S.~Ma, C.~Paganini and B.~F.~Whiting,
%%``Mode stability on the real axis,''
%J. Math. Phys. \textbf{58}, no.7, 072501 (2017)
%doi:10.1063/1.4991656
%[arXiv:1607.02759 [gr-qc]].
%%16 citations counted in INSPIRE as of 22 Oct 2021
%
%%\cite{Leaver:1986gd}
%\bibitem{Leaver:1986gd}
%E.~W.~Leaver,
%%``Spectral decomposition of the perturbation response of the Schwarzschild geometry,''
%Phys. Rev. D \textbf{34}, 384-408 (1986)
%doi:10.1103/PhysRevD.34.384
%%273 citations counted in INSPIRE as of 22 Jun 2020
%
%
%
%%\cite{Lenzi:2021njy}
%\bibitem{Lenzi:2021njy}
%M.~Lenzi and C.~F.~Sopuerta,
%%``Darboux Covariance: A Hidden Symmetry of Perturbed Schwarzschild Black Holes,''
%arXiv:2109.00503 [gr-qc].
%%1 citations counted in INSPIRE as of 04 Dec 2021
%
%%\cite{Chandrasekhar:1976}
%\bibitem{Chandrasekhar:1976}
%S,~Chandrasekhar,
%%``On a transformation of Teukolsky's equation and the electromagnetic perturbations of the Kerr black hole,''
%Proc. R. Soc. Lond. A \textbf{A348}, 39-55 (1976)
%doi:10.1098/rspa.1976.0022.
%
%%\cite{Detweiler:1976zz}
%\bibitem{Detweiler:1976zz}
%S.~L.~Detweiler,
%%``Equations governing electromagnetic perturbations of the Kerr black-hole,''
%Proc. Roy. Soc. Lond. A \textbf{A349}, 217-230 (1976)
%doi:10.1098/rspa.1976.0069
%%18 citations counted in INSPIRE as of 26 Jun 2020
%
%%\cite{Nakajima:2020fhi}
%\bibitem{Nakajima:2020fhi}
%H.~Nakajima and W.~Lin,
%%``Chandrasekhar-like transformations for electromagnetic and scalar waves in Schwarzschild spacetime,''
%Class. Quant. Grav. \textbf{38}, no.2, 027001 (2020)
%doi:10.1088/1361-6382/abc370
%%0 citations counted in INSPIRE as of 22 Oct 2021
%
%%\cite{Hatsuda:2020egs}
%\bibitem{Hatsuda:2020egs}
%Y.~Hatsuda and M.~Kimura,
%%``Semi-analytic expressions for quasinormal modes of slowly rotating Kerr black holes,''
%Phys. Rev. D \textbf{102}, no.4, 044032 (2020)
%doi:10.1103/PhysRevD.102.044032
%[arXiv:2006.15496 [gr-qc]].
%%4 citations counted in INSPIRE as of 25 Oct 2021
%
%%\cite{Bianchi:2021xpr}
%\bibitem{Bianchi:2021xpr}
%M.~Bianchi, D.~Consoli, A.~Grillo and J.~F.~Morales,
%%``QNMs of branes, BHs and fuzzballs from Quantum SW geometries,''
%[arXiv:2105.04245 [hep-th]].
%%6 citations counted in INSPIRE as of 25 Oct 2021
%
%%\cite{daCunha:2021jkm}
%\bibitem{daCunha:2021jkm}
%B.~C.~da Cunha and J.~P.~Cavalcante,
%%``Teukolsky master equation and Painlev\'e transcendents: Numerics and extremal limit,''
%Phys. Rev. D \textbf{104}, no.8, 084051 (2021)
%doi:10.1103/PhysRevD.104.084051
%[arXiv:2105.08790 [hep-th]].
%%3 citations counted in INSPIRE as of 25 Oct 2021
%
%%\cite{Bianchi:2021mft}
%\bibitem{Bianchi:2021mft}
%M.~Bianchi, D.~Consoli, A.~Grillo and J.~F.~Morales,
%%``More on the SW-QNM correspondence,''
%[arXiv:2109.09804 [hep-th]].
%%1 citations counted in INSPIRE as of 25 Oct 2021
%
%%\cite{Kubo:2018cqw}
%\bibitem{Kubo:2018cqw}
%N.~Kubo, S.~Moriyama and T.~Nosaka,
%%``Symmetry Breaking in Quantum geometrys and Super Chern-Simons Matrix Models,''
%JHEP \textbf{01}, 210 (2019)
%doi:10.1007/JHEP01(2019)210
%[arXiv:1811.06048 [hep-th]].
%%9 citations counted in INSPIRE as of 22 Oct 2021
%
%%\cite{Moriyama:2021mux}
%\bibitem{Moriyama:2021mux}
%S.~Moriyama and Y.~Yamada,
%%``Quantum Representation of Affine Weyl Groups and Associated Quantum curves,''
%SIGMA \textbf{17}, 076 (2021)
%doi:10.3842/SIGMA.2021.076
%[arXiv:2104.06661 [math.QA]].
%%1 citations counted in INSPIRE as of 22 Oct 2021
%
%%\cite{Suzuki:1998vy}
%\bibitem{Suzuki:1998vy}
%H.~Suzuki, E.~Takasugi and H.~Umetsu,
%%``Perturbations of Kerr-de Sitter black hole and Heun's equations,''
%Prog. Theor. Phys. \textbf{100}, 491-505 (1998)
%doi:10.1143/PTP.100.491
%[arXiv:gr-qc/9805064 [gr-qc]].
%%88 citations counted in INSPIRE as of 22 Oct 2021
%
%%\cite{Gregory:2021ozs}
%\bibitem{Gregory:2021ozs}
%R.~Gregory, I.~G.~Moss, N.~Oshita and S.~Patrick,
%%``Black hole evaporation in de Sitter space,''
%Class. Quant. Grav. \textbf{38}, no.18, 185005 (2021)
%doi:10.1088/1361-6382/ac1a68
%[arXiv:2103.09862 [gr-qc]].
%%6 citations counted in INSPIRE as of 22 Oct 2021
%
%%\cite{Motohashi:2021zyv}
%\bibitem{Motohashi:2021zyv}
%H.~Motohashi and S.~Noda,
%%``Exact solution for wave scattering from black holes: Formulation,''
%PTEP \textbf{2021}, 083
%doi:10.1093/ptep/ptab097
%[arXiv:2103.10802 [gr-qc]].
%%4 citations counted in INSPIRE as of 22 Oct 2021
%
%
%
%
%%%\cite{Wheeler:1955zz}
%%\bibitem{Wheeler:1955zz}
%%J.~Wheeler,
%%%``Geons,''
%%Phys. Rev. \textbf{97}, 511-536 (1955)
%%doi:10.1103/PhysRev.97.511
%%%458 citations counted in INSPIRE as of 26 Jun 2020
%
%%%\cite{Mashhoon:1973zz}
%%\bibitem{Mashhoon:1973zz}
%%B.~Mashhoon,
%%``Scattering of Electromagnetic Radiation from a Black Hole,''
%%Phys. Rev. D \textbf{7}, 2807-2814 (1973)
%%doi:10.1103/PhysRevD.7.2807
%%%82 citations counted in INSPIRE as of 26 Jun 2020
%
%%%\cite{Chandrasekhar:1980}
%%\bibitem{Chandrasekhar:1980}
%%  S.~Chandrasekhar,
%%  %``On one-dimensional potential barriers having equal reflexion and transmiss%ion coefficients,''
%%  Proc. R. Soc. Lond. A369425-433
%%  doi:10.1098/rspa.1980.0008
%
%%%\cite{Sasaki:1994rw}
%%\bibitem{Sasaki:1994rw}
%%M.~Sasaki,
%%%``PostNewtonian expansion of the ingoing wave Regge-Wheeler function,''
%%Prog. Theor. Phys. \textbf{92}, 17-36 (1994)
%%doi:10.1143/PTP.92.17
%%[arXiv:gr-qc/9402042 [gr-qc]].
%%%55 citations counted in INSPIRE as of 19 Sep 2020



\end{thebibliography}
\end{document}